\documentclass[12pt]{article}  
\newcommand{\x}{x}
\newcommand{\comma}{\;\; ,}
\newcommand{\period}{\;\; .}

\newcommand{\eq}{\; = \;}
\newcommand{\sep}{\;\; , \;\;}
\newcommand{\be}{\begin{equation}}
\newcommand{\bd}{\begin{displaymath}}
\newcommand{\ee}{\end{equation}}
\newcommand{\ed}{\end{displaymath}}
\newcommand{\ba}{\begin{eqnarray}}
\newcommand{\ea}{\end{eqnarray}}
\renewcommand{\L}{N}
\newcommand{\Lc}{L}
\renewcommand{\iota}{i}
\title{Completeness of the Bethe ansatz for the six and eight-vertex models}

\author{ R.J. Baxter\\
{\protect \small Theoretical Physics, I.A.S. and School of Mathematical
Sciences}\\
{\protect \small  The Australian National University,
 Canberra, A.C.T. 0200, Australia  }
}

\date{Thursday 7 February 2002}

\begin{document}


\maketitle

\abstract{We discuss some of the difficulties that have been mentioned in the literature
in connection with the Bethe ansatz for the  six-vertex model and XXZ chain, and for the
eight-vertex model. In particular we discuss the ``beyond the equator'', infinite momenta and
exact complete string problems.  We show how they can be overcome and conclude that the
coordinate Bethe ansatz does indeed give a complete set of states, as expected.}

\vspace{0.5cm}


\section{Introduction}

There are proofs in the literature of the combinatorial completeness of the Bethe
ansatz\cite{Bethe31} for the six-vertex model and XXZ chain:\cite{Kirillov84,Kirillov97} i.e.
that for a lattice of $\L$ columns it gives all $2^{\L}$ eigenvectors (states) of the transfer
matrix. In the presence of an arbitrary  field $H$  there seems  to be no doubt that this is
so, and this is in agreement with the results of the numerical experiments we report in section 2.
Other studies have been made, also indicating the completeness of the Bethe
ansatz.\cite{KlumperZittartz89,Essler91,Essler92,Juett94,Kuniba2000}
   
    Even so, there still appear papers that either question this completeness, or at least
appear to question it, when the field is zero, or at special values of the crossing parameter
$\lambda$ (or $\eta$). Statements have been made that ``the Bethe vector vanishes'' for states
with more down arrows than up arrows,\cite{FaddTak84,Pronko99} and  that it is incomplete or
``singular'' if some of the momenta are infinite.\cite{Sidd98,Wal99,NohLeeKim00} Here we show
that  these problems can be overcome in the coordinate Bethe ansatz if one recognizes that one is
dealing with a set of algebraic equations, so must include an appropriately generalized ``point at
infinity'' in one's considerations, and properly normalize the eigenvector. 

Recently it has been claimed that ``Bethe's equation is incomplete''  at special ``roots of
unity'' values of  $\lambda$.\cite{DegFabMcCoy01} -
\cite{FabMcCoy01c} By this it is meant that the Bethe zeros $v_1, \ldots ,v_n$, or
equivalently the Bethe momenta
$k_1, \ldots ,k_n$, are not uniqely defined. They contain at least one exact complete string,
and one is free to choose each string centre at will.\footnote{Except that if $H=0$, $\L$
is even and $n = \L/2$, then one can require that the eigenvector $g$ also be an
eigenvector of the arrow reversal operator
$R$.  If one does this,  then $v_1,
\ldots ,v_n$ must satisfy the constraint 
(\ref{sixvconstr}).} This freedom is noted in \cite{FabMcCoy01c}, after
equation (1.35) therein. We show that this freedom is because the eigenvalue is degenerate, so 
the eigenvector itself is not uniqely defined. Any allowed choice of
$v_1,\ldots ,v_n$ gives a valid eigenvector. The set of such choices is a curve in the
eigenspace. For the simplest case, which is when $v_1, \ldots ,v_n$ form just one single
string, we show that the vectors on this curve span the eigenspace. Thus the Bethe ansatz
is complete for this case, precisely because of this lack of uniqueness. We fully expect this
argument to generalize to more complicated cases.

In all the cases we have looked at, we have found that the Bethe ansatz is indeed complete: it
gives all the eigenvectors (states). More precisely, it can be used to construct a basis for each
eigenspace. \footnote{This contradicts  the statement in the abstract of \cite{FabMcCoy01a}, and
repeated in the introduction of \cite{Deguchi01a}, that ``the Bethe ansatz equations
determine only the eigenvectors which are the highest weights of the infinite dimensional $sl_2$
loop algebra''. }

Apart from the difficulty mentioned in section 6, the problems we encounter 
can be resolved by the methods mentioned after equation (\ref{sdef2}) and reviewed in the
summary.

We also present the coordinate Bethe ansatz equations
for the eight vertex model in zero field,
with an even number of columns, and discuss how the infinite momenta and exact complete
string problems can be resolved. We expect these equations to be similarly complete. 

We further show that the functional relation between the eigenvalues $T(v), Q(v)$ of the $T$
and $Q$ matrices can itself be written as a generalized eigenvalue problem, with the Fourier
coefficients of $Q(v)$ being the elements of the eigenvector.

In one sense these problems have to be resolvable, since if the ansatz is complete for
arbitrary field $H$ and crossing parameter $\lambda$, then (at least in principle) one can
always deal with difficult cases by taking a limit. (When $\lambda$ is real or pure
imaginary, and the field $H$ is zero or  pure imaginary, then one can choose the
spectral variable $v$ so that the transfer matrix is hermitian, so we know it is
diagonalizable. Since the eigenvectors are independent of $v$, this means that it is
diagonalizable for all $v$.)  However, the question is whether one can do better than this
and find all the eigenvectors for any particular values of $H$ and
$\lambda$. For the problems that have been addressed in the literature, we believe the answer
to be yes.

To paraphrase Mark Kac, this paper is intended to fill a much-needed
gap in the literature.

\section{The six-vertex model in a field }

The six-vertex model \cite{Lieb67a} - \cite{Baxter82} has Boltzmann weights $\omega_1,
\ldots ,\omega_6$. Let 
\bd
R_{++} \eq \left(\begin{array}  {cc} \omega_1 & 0 \\ 0 & \omega_4 \end{array} \right)
\sep
R_{+-} \eq \left(\begin{array}  {cc} 0 & 0 \\ \omega_6 & 0 \end{array} \right)
 \comma \ed
\be
R_{-+} \eq \left(\begin{array}  {cc}  0 & \omega_5 \\ 0 & 0 \end{array}
\right)
\sep
R_{--} \eq \left(\begin{array}  {cc} \omega_3 & 0 \\ 0 & \omega_2 \end{array} \right)
 \comma \ee

Let $\sigma = {\sigma_1, \ldots , \sigma_{\L }}$ denote the state of a row of $\L $ vertical
arrows ($+1$ for an up arrow, $-1$ for a down arrow). Then the transfer matrix for a
lattice of $\L $ rows is the $2^{\L}$ by $2^{\L}$ matrix
$T$ with elements
\be 
T_{\sigma, \sigma'} \eq {\rm Trace } \,  R_{\sigma_1, \sigma_1'} R_{\sigma_2, \sigma_2'} 
\cdots  R_{\sigma_{\L}, \sigma_{\L}'} \period \ee
Considered as a function of $\omega_1, \ldots , \omega_6$, it has the symmetries:
\be  \label{Ttrans}
T(\omega_1,\ldots ,\omega_6 ) = (-1)^{\L} T(-\omega_1,\ldots ,-\omega_6 ) =
T^{t}(\omega_4,\omega_3,\omega_2,\omega_1,\omega_6,\omega_5) \comma \ee
the superfix $t$ denoting transposition.

In statistical mechanics one wants to calculate the partition function
\be Z \eq {\rm Trace } \; T^{M_r} \comma \ee
whrer ${M_r}$ is the number of rows of the lattice. It is therefore desirable to diagonalize 
the matrix $T$. This problem has been solved (for  general values of $\omega_1 ,\ldots ,
\omega_6$ ) by the  Bethe ansatz \cite{Suth67} - \cite{BaxterSIAM71}.\footnote{In fact, 
\cite{BaxterSIAM71} is more general yet, considering an inhomogeneous model where the
field $H$ and the rapidity variable $v$ vary from column to column. Here we shall only
consider the homogeneous model.}

 In the Bethe ansatz one
characterizes the state ${\sigma_1, \ldots , \sigma_{\L}}$  by the positions $X = {x_1, 
\ldots , x_n}$ of the down
arrows, i.e. $\sigma_j = -1$ iff one of $x_1,\ldots ,x_n$ is equal to $j$, else 
$\sigma_j = +1$. Because of the ``ice rule'' that there be two arrows into each vertex, and two
arrows out, the number
$n$ is conserved, being the same on all rows of the lattice. For a lattice of ${\L}$ columns,
the transfer matrix $T$ therefore breaks up into ${\L}+1$ diagonal blocks, one for each
value of $n$ from 0 to ${\L}$. Within block $n$, an eigenvector $g$ has elements $g(X) =
g(x_1,\ldots ,x_n)$ for each state $\sigma$ or  $X$.

The Bethe ansatz is the following guess at the elements of the eigenvector:
\be \label{gexp} g(\x_1,\ldots ,\x_n) = \sum_P A(p_1,\ldots, p_n) \; e^{\iota k_{p_1}{\x_1}}
\cdots e^{\iota k_{p_n}{\x_n}}  \ee
where the sum is over all the $n!$ permutations $\{ p_1,\ldots p_n \}$ of 
$\{ 1,\ldots ,n \}$, and the integers
$\x_1, \ldots, \x_n$ lie in the range
\be \label{range}
1 \leq \x_1 < \x_2 < \cdots < \x_n \leq {\L}  \period \ee

Substituting this ansatz directly into the eigenvalue/eigenvector equations, one finds the
following sufficient conditions  for $g$ to be an eigenvector: 
\be \label{aperm}
s_{p_j,p_{j+1}} A(p_1,\ldots, p_n)
+ s_{p_{j+1},p_{j}} A(p_1,\ldots, p_{j+1},p_j,\ldots ,p_n) = 0 \comma \ee

\be \label{acyc} e^{\iota {\L} k_{p_1}} A(p_2,\ldots ,p_n,p_1) = A(p_1,\ldots ,p_n) \comma \ee
for 
$j=1,\ldots ,n-1$ and all permutations
$\{ p_1,\ldots, p_n \}$.

Here
\be \label{sij} s_{j,m} = \omega_1  \omega_3 - ( \omega_1  \omega_2 + \omega_3  \omega_4 -
\omega_5  \omega_6 )
e^{\iota k_m} +\omega_2  \omega_4 e^{\iota (k_j+k_m)} \period
\ee 

We emphasize that (\ref{gexp}) - (\ref{sij}) are {\em sufficient} conditions (together with $g
\neq 0$) for $g$ to be an eigenvector. They involve the Boltzmann weights only via the ratios
\be \label{ratios}
(\omega_1  \omega_2 + \omega_3  \omega_4 -
\omega_5  \omega_6 )/(\omega_1 \omega_3) \; \; \\{\rm and } \; \; \omega_2 \omega_4/(\omega_1
\omega_3) \period \ee
This implies that the transfer matrices of two models with different weights, but the same
values of these ratios, {\em commute}. This can be proved directly by appropriately extending the
method of section 9.6 of \cite{Baxter82}, or of \cite{Baxter71} or \cite{Baxter72}).

 The eigenvalue corresponding to this eigenvector is
\be \Lambda \eq \omega_1^{\L} \prod_{j=1}^n 
\frac{ \omega_1 \omega_3 + (\omega_5 \omega_6 - \omega_3 \omega_4) e^{\iota
k_j} } { \omega_1(\omega_1 - \omega_4 e^{\iota k_j} )}
 + \omega_4^{\L} \prod_{j=1}^n \frac{ \omega_1 \omega_2  - \omega_5 \omega_6  - \omega_2
\omega_4 e^{\iota k_j} } { \omega_4(\omega_1 - \omega_4 e^{\iota k_j}) } \period \ee

If the weights $\omega_1 , \ldots , \omega_6$ are all non-zero, it is convenient to write 
them as 
\bd \label{weights2}
  \omega_1 = e^{H+V} a \sep \omega_2 = e^{-H-V} a  \sep \omega_3 = e^{H-V} b
\ed
\be  \omega_4 = e^{-H+V} b \sep  \omega_5 = \omega_6 = c \period \ee
Here $H$ is a dimensionless electric field in the horizontal direction; $V$ is the field in
the vertical direction.\footnote{The weights $\omega_5, \omega_6$
are sinks and sources of horizontal arrows, so only occur in the partition function and
transfer matrix in the product combination $\omega_5
\omega_6$. This means that there is no loss of generality in choosing 
$\omega_5 =  \omega_6 = c$.}
 
Defining
\be     \Delta  = (a^2+b^2-c^2)/(2 a b) \sep e^{\iota {\kappa}_j}  = e^{-2H} e^{\iota k_j}
\comma
\ee and dividing $s_{jm}$ by a constant factor $e^{2 H} a b$ that cancels out of
(\ref{aperm}),  the above equations involving 
$\omega_1 \ldots ,  \omega_6$ simplify to 
\be \label{defsjm}
s_{j,m} \eq 1 - 2 \Delta e^{\iota {\kappa}_m} + e^{\iota ({\kappa}_j + {\kappa}_m )}
\comma \ee

\be \label{eigval}
\Lambda \eq e^{({\L}-2n)V} \left\{ e^{{\L}H} \prod_{j=1}^n \frac{ a b  + (c^2 - b^2) e^{\iota
{\kappa}_j} } { a (a - b \, e^{\iota {\kappa}_j} ) } + e^{-{\L}H} \prod_{j=1}^n
 \frac{ a^2   - c^2  - a b  \, e^{\iota {\kappa}_j} } {
 b (a - b \, e^{\iota {\kappa}_j)} } \right\} \period \ee

Note that $V$ enters the above equations only via a factor $e^{({\L}-2n)V}$ in the eigenvalue
$\Lambda$. This is because of the ice rule: $T$ commutes with the diagonal matrix $S_z$ 
which has  entries $(\sigma_1 + \sigma_2 + \cdots + \sigma_{\L})
\delta_{\sigma, \sigma'}$. The effect of $V$ in the Boltzmann weights (\ref{weights2}) is to
pre-  and post-multiply $T$ by $\exp (S_z V/2 )$. Within the diagonal block $n$, the diagonal
entries of $S_z$ are
 ${\L}-2n$, so the effect of introducing $V$  is to merely multiply  all entries of $T$  by
$e^{({\L}-2n)V}$. Without loss of
generality we shall from now on take $V = 0$.

The next step is to derive what Fabricius and McCoy call ``Bethe's equation''. \cite[eqn.
1.2]{FabMcCoy01a} This is
easily done when the $s_{j,m}$ are all non-zero, but much of the apparent confusion in
the literature arises when some of them are zero, so we proceed
carefully. To be an eigenvector,
$g$ must be non-zero, so at least one of the coefficients 
$A(p_1,\ldots, p_n)$ must be non-zero. In Appendix A we show that this implies that
\be \label{BA} e^{\iota {\L} k_j } \, \prod_{ m=1 , m \neq j} ^n s_{j,m} \eq 
(-1)^{n-1} \prod_{ m=1 , m \neq j } ^n s_{m,j} \ee 
for $j = 1, \ldots , n$.  Hence these $n$ equations are a necessary consequence of 
the linear equations (\ref{aperm}) and (\ref{acyc}) for the $n!$ coefficients $A(p_1,\ldots, p_n)
$. We discuss below their sufficiency for the case when all of the
$s_{j,m}$ are  non-zero.

Also, substituting all $n$ cyclic permutations of $\{ p_1, \ldots, p_n \} $ into (\ref{acyc}),
one gets $n$ linear homogeneous equations for the $n$ coefficients $A$ that occur. Since at
least one of them is non-zero (for some permutation $\{ p_1, \ldots, p_n \} $), the determinant of
coefficients must vanish, giving 
\be \label{sumks}
e^{\iota {\L}( k_1 + k_2 + \cdots + k_n)} = 1 \period \ee

We refer to (\ref{gexp}) - (\ref{sumks}) as ``the Bethe ansatz equations''.

\subsubsection*{Transformation to difference variables}

Define $\rho, \lambda, v$ so that
\ba \label{parmdiff}
a & = & \rho \sinh [(\lambda-v)/2] \nonumber \\
b & = & \rho \sinh [(\lambda+v)/2]  \\
c & = & \rho \sinh \lambda \nonumber \ea
then
\be \label{Delta} \Delta = - \cosh \lambda \ee

Define also $v_1, \ldots , v_n$ so that
\be \label{eik} e^{\iota k_j} \eq e^{2 H} e^{\iota \kappa_j} \eq e^{2 H} \frac{e^{\lambda} - e^{v_j} }
{e^{\lambda + v_j} -1 }
\period \ee Then
\be \label{defs2}
s_{i,j} = \frac{ \sinh  \lambda \; \sinh[(v_i - v_j + 2 \lambda )/2 ]}
{\sinh[(v_i + \lambda)/2] \, \sinh[(v_j + \lambda)/2] } \period \ee

If we define functions $\phi (v) , Q(v)$ by
\bd \phi (v)  = \rho^{\L} \sinh ^{\L} (v/2) \comma \ed
\be \label{qdef}
Q	(v) = \prod_{j=1}^n \sinh [(v-v_j)/2] \comma \ee
then (\ref{eigval}) can be written in the form
\be \label{eig2}
\Lambda \eq (-1)^n \; \frac{e^{{\L}H} \, \phi (\lambda-v)\, Q(v + 2 \lambda) +
e^{-{\L}H} \, \phi (\lambda+v) \, Q(v - 2 \lambda) }{ Q(v)} \period \ee

We shall also introduce a parameter $q$ by
\be \label{defq}
q \eq - e^{\lambda} \sep  \Delta = (q + q^{-1} )/2 \period \ee
This $q$ is the $-q$ of \cite{FabMcCoy01a}.

\subsubsection*{Commuting transfer matrices}  

In the notation (\ref{parmdiff}) - (\ref{Delta}), the ratios (\ref{ratios}) that enter the
eigenvector calculation are
\bd
2 \, e^{-2H}  \Delta \; \; {\rm and } \; \;   e^{-4H} \period \ed

Keeping $\rho, \lambda , H$ fixed, and writing the transfer matrix $T$ as a  function $T(v)$ of
$v$, it follows that 
\be \label{Tcomm}
T(v) T(v' ) \eq  T(v') T(v )  \ee
for all $v, v'$. The transfer matrices $T(v)$, $T(v')$ {\em commute}.

Further, if we define the Pauli matrices
\be 
\sigma_j^x = \left( \begin{array} {cc} 0 & 1 \\ 1 & 0 \end{array}  \right) \sep
\sigma_j^y = \left( \begin{array} {cc} 0 & -i \\ i & 0 \end{array} \right) \sep 
\sigma_j^z = \left( \begin{array} {cc} 1 & 0 \\ 0 & -1 \end{array} \right) \ee
acting on the spin (arrow) in position $j$,  then the logaritmic
derivative of  $T(v)$ at
$v = -\lambda$ is  a linear combination of the identity operator and the hamiltonian
\bd
{\cal H} = - \frac{1}{2} \, \sum_{j=1}^{\L} \! \left\{
\cosh H \left( \sigma_j^x \sigma_{j+1}^x \! +  \! \sigma_j^y
\sigma_{j+1}^y   \right)  \! - \! i \, \sinh H \left( \sigma_j^x \sigma_{j+1}^y \! - \!
\sigma_j^y
\sigma_{j+1}^x \right) \! + \! 
 \Delta \, \sigma_j^z \sigma_{j+1}^z   \right\} \ed
interpreting suffixes ${\L}+1$ as $1$. Hence $T(v)$ also commutes with this hamiltonian.

Incrementing $v$ by
$2 \pi i$ is the same as negating all of $a, b, c$.\footnote{Remember that $\omega_5$ and
$\omega_6$ always occur in pairs, so negating $c$ has no effect on $T$.} This merely multiplies
$T$ by $(-1)^{\L}$. The lowest and highest powers of $e^{v/2}$ that can occur in the
expansion of an element of $T(v)$ are $-{\L}$ and ${\L}$. It follows that $T(v)$ can be
expanded in the form
\be \label{Texpn}
T(v) \eq \sum_{r=0}^{\L} T_r e^{({\L}-2r)v/2} \comma \ee
the coefficients $T_r$ being matrices independent of $v$.

The commutation relations (\ref{Tcomm}) imply that $T_0, \ldots , T_{\L}$ all commute with
one another, and with $\cal H$.

\subsubsection*{Hermiticity of $T$ and unitarity of $P$}

Negating $v$ and $H$ interchanges $\omega_1$ with $\omega_4$, and $\omega_2$ with 
$\omega_3$. From (\ref{Ttrans}), this transposes the transfer matrix $T$. If $\lambda, \rho$ are
real and $H, v$ are pure imaginary, or if  $v$ is
real and $\rho, H, \lambda$ are pure imaginary, this implies that $T$ is hermitian. Thus it is
diagonalizable. Combining this with the commuation properties above, it follows that there
exists a  unitary eigenvector  matrix  $P$ such that
\be \label{Pdiag}
P^{\dagger} T_r P = \; \; {\rm and} \; \;  P^{\dagger} {\cal H} P = \; \; {\rm diagonal} \period
\ee

In both cases  $\Delta$ is real. 
For these cases, the eigenvalue $\Lambda$ must be  real. Since $H, v/\lambda$ are pure
imaginary in both, (\ref{eig2}) suggests that $Q(v+2 \lambda)^* = Q(v-2 \lambda)$.
 From (\ref{qdef}), this implies that:

(i)~~ $\lambda$ real: $v_1, \ldots ,v_n$ are either pure imaginary or occur in pairs
$v_j, -v_j^*$ positioned symmetrically about the imaginary axis;

(i)~~ $\lambda$ pure imaginary: $v_1, \ldots ,v_n$ are either real or occur
in complex conjugate pair.

In both cases it follows that the wave numbers $k_1, \ldots , k_n$ are either real or
occur in complex conjugate pairs. Similarly for $\kappa_1 , \ldots , \kappa_n$.

Note that $P$ is independent of $v$, so provided $H$ is pure imaginary, we have
provedthat $T = T(v)$ is diagonalizable and $P$ is unitary for {\em all} complex  $v$.

Of course, we usually take the Boltzmann weights $\omega_1, \ldots , \omega_6$ to be positive
real, so from that point of view we would like to take $H$ to be real, rather than pure
imaginary. However, it is reassuring to have a proof that $T$ is diagonalizable even if the
proof only holds when $H$ is pure imaginary. This does include the ``zero-field'' case of
main interest, as well as other problems thta have been looked at, such as the critical
Potts model with $q < 4$.\footnote{In section 6.2 of \cite{BaxKellWu76} it is shown that this
model is equivalent to a six-vertex model with a boundary seam. This seam is
equivalent to introducing a horizontal field $H = \theta/{\L}$, where $q^{1/2} = 2 \cosh
\theta$, so $\theta$ and $H$ are pure imaginary when $q < 4 $ and the model is critical.}

This  implies that $T$ will be diagonalizable for all
$H$ except posssibly for isolated non-zero values off the imaginary axis. Our question is:
does the Bethe ansatz give all the eigenvectors?

\subsubsection*{The matrix $\tilde{Q} (v)$} 

Each eigenvalue $\Lambda$ must have an expansion corresponding to (\ref{Texpn}) :

\begin{equation} \label{BAalt1} \Lambda(v) = \sum_{r=0}^{\L} t_r e^{({\L}-2r) v/2}  
\end{equation}
  Thus $\Lambda$ is an entire function of $v$ and the RHS of (\ref{eig2}) must
vanish when the denominator does, giving 
\be \label{pheqns}
e^{{\L} H} \phi (\lambda-v_j) \, Q(v_j + 2 \lambda) + e^{-{\L} H}
\phi (\lambda+v_j) \, Q(v_j - 2 \lambda) = 0 \comma \ee
for $j = 1, \ldots ,n$.

These are precisely the equations (\ref{BA}). The author noted (for H=0) in 1971 that they imply
the existence of a matrix $\tilde{Q}(v)$ with eigenvalues $Q(v)$ that commutes with $T(v)$ and
satisfies the matrix functional relation
\be \label{fnlreln}
T(v) \tilde{Q} (v)  \eq  e^{{\L} H} \phi (\lambda-v)\, \tilde{Q}(v + 2 \lambda') +
e^{- {\L} H} \phi (\lambda+v) \, \tilde{Q}(v - 2 \lambda') \comma \ee
where $\lambda' = \lambda - i \pi$. This proved to be the key to solving 
the eight-vertex model.\cite{Baxter71, Baxter72} 

For the zero-field eight-vertex model  an explicit 
construction for $\tilde{Q}(v)$ is given in 
 section 6 of \cite{Baxter73a} and in section (10.5)  of
\cite{Baxter82}.\footnote{As we remark in section 7, there is
a typing error in eqn (10.5.8) of \cite{Baxter82}: $\sigma_{j+1}$ therein should be
$\sigma_{j-1}$.} This is specialized to the six-vertex model in equations (8), (96) and (97) of
\cite{Baxter73a}, for $H=0$ and ${\L}$ even.  

 The vectors $\Phi_n(v|\sigma )$ therein are generalizations of the special eigenvectors $\psi$
we discuss in section 3. They form the columns of a matrix $\Phi_n(v )$. Writing
$[N,m] = N!/(m! (N-m)!)$, this matrix  has $[{\L},n]$ rows and
$[{\L},{\L}/2]$ columns. Provided the rows are linearly independent, we can define the 
$[{\L},n]$ by $[{\L},n]$  matrix $\tilde{Q}(v)$ by
\be \label{cnstrctQ}
{\tilde Q}(v) \Phi_n(v_0|\sigma ) = \Phi_n(v|\sigma ) \ee
 $v_0$ being an arbirary fixed parameter,  for all $n$ from 0 to ${\L}$. Then it is shown
in \cite{Baxter73a}) and \cite{Baxter82} that
\be
{\tilde Q}(v) T(v) = T(v) {\tilde Q}(v) \comma \ee
so we can simultaneously diagonalize both $T(v)$ and  $\tilde{Q}(v)$.
Doing this, we find (for $H=0$ and $\L $ even) that the eigenvalues $Q(v)$ must indeed have the
form given in equation (\ref{Qexp}) below.

In the cases discussed in section 5, where some of $v_1, \ldots ,v_n$
form one or more complete strings, the eigenvalue $\Lambda$ of $T(v)$ is degenerate. This is
reflected in the fact that the string centres (the average value of the $v_j$ within a string)
are not determined by the Bethe ansatz. One of the main objects of this paper is to stress that
this is not a deficiency or ``incompleteness'' of the Bethe ansatz, but rather a strength. It
means that it can be used to construct a complete basis of the eigenspace.

However, the corresponding eigenvalues of ${\tilde Q}(v)$ are {\em not} degenerate, so one can 
fix the string centres by requiring that $g$ in (\ref{gexp}) be also an eigenvector of 
${\tilde Q}(v)$. This is what Fabricius and McCoy have achieved. In physicist's terms, they
have resolved the degeneracy of the eigenvalue $\Lambda$; in mathematician's terms, they have
made a particular choice of the basis of the eigenspace of $T(v)$.

\subsubsection*{Beyond the equator: the relation between the $n$ and ${\L}-n$ \\ solutions } 

As we note below, there appears to be no problem solving these equations in the
presence of a non-zero field
$H$ (pure imaginary or real), even when $n > {\L}/2$ and there are more down arrows than up.
However, the eignvalues $\Lambda$ are then the same as the mirror case (where all arrows
are reversed), with
$n \rightarrow {\L}-n$ and $H
\rightarrow -H$. Is there a relation between the two solutions?

This problem has been studied by Bazhanov, Lukyanov and
Zamolodchikov \cite{BazLukZam97,BazLukZam99}. Consider
the equation (\ref{eig2}) for some $n$ and $H$, with a function $Q_1(v) = Q(v)$,  and
the same equation with $n, H$ replaced by ${\L}-n, -H$, with a different function $Q_2(v) = Q(v)$
but the same $\Lambda$. Eliminate $\Lambda$  between the two equations. We obtain
\be \label{Wrper1}
W(v+\lambda) \eq W(v-\lambda) \comma \ee
where the ``Wronskian'' $W(v)$ is defined by
\be W(v) \eq \frac{e^{{\L}H} Q_1(v+\lambda) Q_2 (v-\lambda)  - (-1)^{\L} e^{-{\L} H} Q_1(v-\lambda) Q_2
(v+\lambda)  }{\phi(v)} \period \ee

We continue to require that $Q_1 (v)$ and $Q_ 2(v)$ be of the form (\ref{qdef}), with $n$
replaced by $n$, ${\L}-n$, respectively. It follows that
\be\label{Wrper2}  W(v+ 2 \pi i) \eq W(v)  \period \ee

For arbitrary $\lambda$, excluding the ``root of unity'' cases discussed in section 5 in which
$i \lambda$ is a rational fraction of $\pi$, the only solution of both (\ref{Wrper1}) and
(\ref{Wrper2}) is that
$W(v)$ be a  constant
$D$. Hence for non-``root of unity'' cases, $Q_1 (v)$ (with parameters $n, H$), and $Q_2 (v)$
(with parameters ${\L} - n, -H$) must
satisfy the relation
\be \label{Q1Q2}
e^{{\L}H} Q_1(v+\lambda) Q_2 (v-\lambda)  - (-1)^{\L} e^{-{\L} H} Q_1(v-\lambda) Q_2
(v+\lambda)   \eq  D \phi(v) \period \ee

For $H = 0$, this is equation (47) of (\cite{Pronko99}). We shall discuss it further
below, particularly for the case when $H = 0$ and ${\L}$ is even.

 Pronko and Stroganov\cite{Pronko99}  have also addressed this problem, in a  different manner.
Using their terminology, we shall show in section 4 that it is indeed interesting, and perfectly
possible, to consider Bethe's equations `on the wrong side of the equator'.

\subsubsection*{Bethe's equation as a generalized eigenvalue problem } 

We can write (\ref{eig2}) as

\begin{equation} \label{LamQ} \Lambda(v) Q(v) =  (-1)^n \{ e^{{\L} H} \phi(\lambda - v) Q(v+2
\lambda) + e^{- {\L} H} \phi(\lambda + v) Q(v-2 \lambda) \}   \end{equation}
and $ \phi(v),
Q(v)$ as 

\begin{equation} \phi(v) = \sum_{r=0}^{\L} f_r e^{({\L}-2r) v/2}   \end{equation}

\begin{equation} \label{Qexp} Q(v) = \sum_{j=0}^n q_j e^{(n-2j) v/2}   \end{equation}
Substituting these expansions, together with (\ref{BAalt1} ), into  the $\Lambda, Q$
equation above, one obtains

\begin{equation} \label{bqeqn} \sum_{j=0}^n b_{i,j} q_j = 0 , \end{equation}
where  if $0 \leq i-j \leq {\L} $,

\begin{equation} \label{BAalt2} b_{i,j} = -t_{i-j} + (-1)^n\, e^{\lambda {\L}/2} \, [(-1)^{\L}
e^{{\L} H} e^{\lambda (i-3 j + n  - {\L})} + e^{-{\L} H}  e^{\lambda((3 j -i-n }] f_{i-j} , \end{equation}
else $b_{i,j} = 0$. Here $i = 0, \ldots ,{\L}+n$.

We see that we have ${\L}+n+1$ equations for the ${\L}+n+1$ unknowns $t_0, \ldots t_{\L}$ and $ q_0: q_1 :
\cdots : q_n$. These are an alternative form of Bethe's equations. They define the eigenvalue
$\Lambda (v)$  and (usually) the function $Q(v)$. They are linear in the
$t_j$, and homogeneous and linear in the $q_j$.  The $t_j$ play the role of a set of
``eigenvalues'',
$q_j$ that of an ``eigenvector''.

This form of Bethe's equation  has  some advantages which we shall
mention as we come to them in the following four sections. In particular, suppose that we
actually know, or have guessed, the eigenvalue $\Lambda (v)$, and hence  $t_0, \ldots , t_{\L}$. Then
the equations  are a set of homogeneous linear equations for $q_0, \ldots , q_{\L}$, and can be
solved by the standard apparatus of linear algebra. Let ${\bf B}$ be the ${\L} + n + 1$ by $ n + 1$
matrix with elements $b_{ij}$. Then we can distinguish three cases:

1) ${\bf B}$ has rank $n+1$: then there are no solutions for $q_0,\ldots , q_n$. $\Lambda(v)$ is
not an eigenvalue.

2) ${\bf B}$ has rank $n$: there is one  solution for the ratios $q_0: q_1 : \cdots : q_n$. This
presumably means that the eigenvector $g$ is unique: then  $\Lambda(v)$ is an eigenvalue with
degeneracy one.

3) ${\bf B}$ has rank less than $n$: there is more than one  solution for $q_0: q_1 : \cdots :
q_n$. The eigenvector  $g$ is presumably not unique: then $\Lambda(v)$ is an eigenvalue with
degeneracy greater than one.

Thus we can use these simple considerations to determine whether a given eigenvalue is single or
multiple.

\subsection*{``Bethe's equations'' }

There are a huge number of equations in (\ref{aperm}):
$(n-1) \times n!/2$ homogeneous linear equations for the  $n!$ coefficients
$A(p_1,\ldots ,p_n)$. Fortunately  it seems that they always permit at least one
(possible more - this is the source of some of the misunderstandings in the literature)
non-identically zero solution. This is of the form
\be \label{Asoln}
A(p_1,\ldots ,p_n) = \epsilon_P \; C^{-1} \prod_{ 1 \leq i < j \leq n} t_{p_j,p_i} \comma \ee
where $\epsilon_P = \pm 1$ is the sign of the permutation and the $t_{ij}$ must satisfy
\be \label{st}
t_{ij} \, s_{ji} = t_{ji} \, s_{ij} \period \ee
At least one of $t_{ij}$ and $t_{ji}$ must be non-zero, else the $A(p_1,\ldots ,p_n)$ would
all vanish and $g$ would be the zero vector.

If the $s_{ij}$ are all finite and non-zero and the $k_j$ are finite, then we can take $t_{ij}
= s_{ij}$ and choose the normalization factor $C$ to be unity. The problems discussed in this
paper arise when this is not so. (Apart from the equal $v_j$ difficulty touched on in
section 6.) For these cases one should choose $C$ so that the maximum term in the summand in
(\ref{gexp}) is finite and non-zero (say unity). This maximum is to be taken over all
permutations $P$ and all values of $x_1, \ldots x_n$ allowed by (\ref{range}). 

This is the solution given in (8.4.10) of
\cite{Baxter82}, except that there we took $t_{ij} = s_{ij}$ for all $i, j$. We wish to be
more general here so as to cope with the situation  when
some of the $s_{ij} $ vanish.

Substituting (\ref{Asoln}) into (\ref{acyc}), we obtain the  $n$ equations 
\be \label{BA2} e^{i {\L} k_j }  \eq 
(-1)^{n-1} \prod_{ m=1 , m \neq j } ^n t_{m,j} / t_{j,m} \ee 
for $j = 1, \ldots , n$. 
Together with (\ref{st}), this implies (\ref{BA}).

We remind the reader of equation (\ref{defsjm}), namely
\be \label{sdef2}
s_{ij} = 1 - 2 \Delta e^{\iota k_i - 2 H} + e^{\iota (k_i+k_j) - 4 H } \period \ee

We refer to (\ref{st}) - (\ref{sdef2}) as ``Bethe's equations'', which is a slight extension
of the terminology of Fabricius and McCoy \cite[eqn. 1.2]{FabMcCoy01a}. They are {\em
sufficient} conditions for  (\ref{aperm}) and (\ref{acyc}) to have a non-zero solution for
the coefficients $A(p_1,\ldots ,p_n)$. They form a set of {\em coupled } equations for the
$e^{\iota k_j}, s_{ij}$ and the ratios $t_{ij} : t_{ji}$.

Apart from section 6, the problems we shall be discussing occur when some of these variables
are zero or infinite. The resolution is always to re-express (\ref{gexp}) and (\ref{Asoln})
-  (\ref{BA2}) in terms of  finite combinations of powers of the variables $e^{\iota
k_j}$ and $t_{ij}$, to solve the equations simultaneously for these, and to allow for the
possibility of a solution containing one or more arbitrary degrees of freedom.

\subsubsection{Momentum}

When $v = -\lambda$, $e^{(2n-{\L})H} T(v)/c^{\L}$ is the matrix with entries 
$\prod_{i} \delta(
\sigma_i, \sigma_{i+1}' ) $. It shifts all arrows one column to the right. Doing this to the
eigenvector $g$ is equivalent to multiplying $g$ by $\exp[i(k_1 + \cdots + k_n)]$.
Similarly, the matrix  $e^{({\L}-2n)H} T(\lambda)/c^{\L}$ shifts all arrows one to the left. 
Hence, writing $\Lambda$
as $\Lambda (v)$, 
\bd e^{(2n-{\L})H}  \Lambda(-\lambda) /c^{\L} = c^{\L} e^{(2n-{\L})H}/ \Lambda(\lambda) =
e^{i(k_1 + \cdots + k_n)}   \ed
and $e^{i(k_1 + \cdots +
k_n)}$ must be an ${\L}$th root of unity, in agreement with (\ref{sumks}).
From (\ref{eig2}) it follows that 
\be
e^{i(k_1 + \cdots + k_n)}  \eq  e^{2 n H} Q(\lambda)/Q(-\lambda) \period \ee

\section*{Numerical calculations} 

\subsubsection*{Method} 

To fix our ideas, we conducted a  number of numerical experiments on the
above equations for lattices of small size: up to $ {\L} = 8$. We fixed $H$, $\lambda$ and $\rho$
and evaluated the matrix coefficients $T_r$ in (\ref{Texpn}). We then assigned $v$ an arbitrary
value and diagonalized $T(v)$ directly to obtain the eigenvector matrix $P$, and verified that
it did indeed diagonalize all the $T_r$. For each eigenvalue $\Lambda$ this
gave us the coefficients $t_r$ in (\ref{BAalt1}). We constructed the matrix $\bf B$ and
determined its null space, and hence all solutions of (\ref{bqeqn}) for the $q_j$. This
gave us the function $Q(v)$. We calculated its zeros to obtain the $v_j$ from
(\ref{qdef}). We then calculated the $\kappa_j$ and the wave numbers $k_j$ from
(\ref{eik}), and then the $s_{ij}$ from (\ref{sdef2}).
 Usually the $s_{ij}$ were {\em all} non-zero and we took $t_{ij} = s_{ij}$, and
calculated the coefficients
$A(p_1, \ldots ,p_{\L})$ from (\ref{Asoln}) and the elements of $g$ from (\ref{gexp}). Finally we
normalized this vector (so its largest element was one) and compared it with the correspondingly
normalized column of the matrix $P$.

\subsubsection*{Results for $H \neq 0 $} 

We first took $H$ to be non-zero, either real or pure imaginary, and $\lambda$ to also be either 
 real or pure imaginary. At this stage we avoided the ``roots of unity'' cases when $i \lambda$
is a rational fraction of $\pi$. We encountered no problems with the above procedure.
For a given number $n$ of down arrows, we found no  eigenvalues of $T(v)$ that were
identically degenerate for all $v$.\footnote{Obviously they can be degenerate for
special values of $v$: if $v = \pm \lambda$, then $c^{-{\L}} T(v)$ is the ``momentum''operator
that shifts all arrows in a row one column to the right (or one to the left) and has
eigenvalues which are ${\L}$th roots of unity.}
The column nullity of
$\bf B$ was always 1, so there was only one solution (to within normalization) of
(\ref{bqeqn}) for the
$q_j$. The 
$s_{ij}$ were all non-zero and the wave numbers $k_i$ were all distinct so that (\ref{gexp})
gave a unique non-zero eigenvector. We worked to about 17 decimal digits of accuracy, and the
error in the eigenvector elements was no bigger than $10^{-13}$.

For $H$ pure imaginary, we also observed that the eigenvector matrix $P$ was unitary, that
for every  $v_j$ there was a conjugate according to the rule given after eqn.  (\ref{Pdiag}), and
that the wave numbers $k_1, \ldots , k_n$ were either real or occurred in complex conjugate
pairs. The functions $Q_1(v), Q_2(v)$ had $n, {\L}-n$ finite zeros respectively, and satisfied 
(\ref{Q1Q2}).

\section{Particular values of $H, \lambda$: some very special eigenvectors}

Here we consider the case when $H, \lambda$ satisfy the relation
\be \label{condinit}
\left( - e^{ \lambda \pm 2H} \right)^{\L} \eq 1 \comma \ee
and show that the Bethe ansatz then admits some some very special eigenvectors. 
They are the analogues of the special eigenvectors of the
zero-field eight-vertex  model obtained by the author in section 7 of \cite{Baxter73a}.

Almost the simplest ansatz one can imagine for an eigenvector 
$\psi$ of the $2^{\L}$ by $2^{\L}$ transfer matrix
$T(v)$ is the direct product form
\be  \psi \eq \left( \begin{array} {c} 1 \\ g_1 \end{array} \right) \otimes 
\left( \begin{array} {c} 1 \\ g_2 \end{array} \right) \otimes \cdots \otimes
\left( \begin{array} {c} 1 \\ g_{\L} \end{array} \right) \comma \ee
where $g_1, \ldots, g_{\L}$ are some parameters to be determined.

Define $q$  by (\ref{defq}): $q = - e^{\lambda}$, and let 
\be {\tilde{q}} = q^{\pm 1} \comma \ee
making one of the two possible sign choices here and in the following equations. 

Following the 
method  of section 9.8 of \cite{Baxter82}, we find that $T(v) \psi$ has a simple structure if
\be
g_j = (e^{2H} {\tilde{q}} )^{ j} \, g \; \; {\rm for}\; \; j = 1, \ldots , {\L} \comma \ee
except that we need the cyclic boundary condition $g_{\L + 1} = g_1$: this implies
\be \label{cond1}
{( e^{2H} {\tilde{q}}   )}^{\L} = 1 
\comma \ee
which is (\ref{condinit}). The parameter $g$ is arbitrary: we can choose it at will. Hence we can
regard $\psi$ as a function $\psi (g)$ of $g$.

Then we find that
\be \label{Tg}
T(v) \psi(g) \eq \omega_1^{\L} \psi(g') + \omega_4^{\L} \psi(g'') \comma \ee
 where
\be
g' = {\tilde{q}}  g \sep g'' = {\tilde{q}}^{-1}  g \period \ee

Now look at the sub-space with $n$  down arrows and $S_z = \L - 2 n$. Then
$\psi (g) = g^n \phi_n$, where, in terms of the
positions 
$x_1, \ldots, x_n$ of the down arrows, $\phi_n$ is a vector with entries
\be \label{ddphi}
\phi_n (x_1, \ldots, x_n ) \eq ( {\tilde{q}}  )^{ ( x_1 + \cdots + x_n)}
\period \ee

In this sub-space
(\ref{Tg}) becomes
\be T(v) \phi_n \eq \left( a^{\L} e^{\L H} {\tilde{q}}^{ n} + 
b^{\L} e^{ - \L H} {\tilde{q}}^{- n} \right) \phi_n \period \ee

Thus $\phi_n$ is an eigenvector of $T(v)$, with eigenvalue
\be \label{egspec}
\Lambda \eq   a^{\L} e^{\L H} {{\tilde{q}} }^{ n} + 
b^{\L} e^{ - \L H} {{\tilde{q}} }^{- n} \period \ee
This is true for all $n$ from 0 to $\L$, provided only that the restriction
(\ref{condinit}), or more specifically (\ref{cond1}), is satisfied.

\subsubsection*{Reconciliation with the Bethe ansatz}

How can we reconcile this with the Bethe ansatz? Simply by taking
\be
v_j \rightarrow \mp \, \infty \; \; {\rm for} \; \; j = 1 ,\ldots , n \period \ee
Then, from (\ref{eik}),
\be \label{eikspec}
e^{\iota k_j } \rightarrow  e^{2H} {\tilde{q}}  \; \; {\rm for} \; \; j = 1 ,\ldots , n \comma
\ee so all the exponential factors containing $x_1 ,\ldots ,x_n$ in (\ref{gexp}) are
proportional to $\phi_n (x_1, \ldots, x_n )$ and
\be
g \propto \phi_n \period \ee

Also, from (\ref{qdef}) and (\ref{eig2}), we obtain the result 
(\ref{egspec}) for the eigenvalue $\Lambda$.

It would seem that there is nothing more to say: the vector $\phi_n$ is a special case of
the Bethe ansatz when all the $v_j$ tend to $\mp \infty$. The eigenvalue is given by
(\ref{eig2}).

However,  the alert reader will notice that we have said nothing about Bethe's equations. More
seriously, all the $k_j$ are finite and equal. This is a potential problem in itself, since
a casual inspection of (\ref{st}) suggests that it implies that $t_{i,j} = t_{j,i}$ for all
$i,j $ from 1 to $n$. If we then take $A(P)$ to be given by (\ref{Asoln}) and substitute into
(\ref{gexp}), all terms will be equal except for the sign factor $\epsilon_P$. They will
therefore all cancel and we shall obtain $g=0$. (We return to the general problem of what happens
when $k_i = k_j$, $s_{i,j} = s_{j,i} \neq 0 $ for some pair of values
$i,j$, in section 6.)

However, $e^{i\kappa_j} = {\tilde{q}} $, so from (\ref{defsjm}) and  (\ref{defq}),
\be s_{j,m} = 0 \; \; {\rm for \; \, all} \; \; j,m \period \ee

At first sight this appears to only make matters worse. If we use the ``normal'' solution  of
(\ref{st}), namely $t_{j,m} = s_{j,m}$, then (\ref{Asoln}) gives $A(P) = 0$. Now each term in
(\ref{gexp}) vanishes, not just their sum!

The answer is of course that one should not use this solution. In fact there is now no reason to
use the subsidiary ``ansatz''  (\ref{Asoln}) at all. For any finite choice of coefficients $A(P)$,
the equations (\ref{aperm}) are satisfied simply because the $s_{j,m}$ vanish. All that remains is
to satisfy (\ref{acyc}). One simple choice that does this is to take
 \be A(P) = 1 \ee
for all permutations $P$.
From (\ref{cond1}) and  (\ref{eikspec}),
\be
e^{\iota \L k_j}  = 1 \comma \ee
so both (\ref{aperm}) and (\ref{acyc}) are satisfied  and
 (\ref{gexp}) becomes
\be g(x_1, \ldots x_n) \eq n! \, e^{\iota k (x_1 + \cdots + x_n)} \comma \ee
where $k = k_1 = \cdots = k_n$. Apart from the non-zero normalization factor $n!$ , this is
the result (\ref{ddphi}).

These special eigenvectors are a good illustration of how one can satisfy the original Bethe ansatz
equations when some of the $s_{i,j}$ vanish. {\em The form (\ref{Asoln}) is not part of the
original ansatz, but an addition to it.} It is a necessary consequence of (\ref{aperm}) if
all of the
$s_{i,j}$ are non-zero, but if enough of them vanish it ceases to be necessary.

Still, if (\ref{Asoln}) is true in general we should not be too ready to
abandon it in particular. We can still take $A(P)$ to be given by (\ref{Asoln}). Equation
(\ref{st}) now imposes no restriction on the $t_{i,j}$, but (\ref{BA2}) does. For the case that we
are discussing in this section, a simple solution of (\ref{BA2}) that avoids the problems that
occur when $t_{i,j} = t_{j,i}$ is to take $t_{i,j} = - t_{j,i} \neq 0 $ for all $i,j$. Then $A(P)$
is independent of $P$ and we regain the solution just discussed.

Another solution can be obtained, under the more specialized conditions (\ref{Hzero}) and
(\ref{constr}), by letting the $v_j$ in section 5, for the case $M = n$,  tend to $ \mp \,
\infty$. 

More generally, we can take the $t_{ij}$ to be arbitrary and non-zero, with $t_{ij} \neq t_{ji}$,for
$1 \leq i, j
\leq n-1$, and then use (\ref{BA2}) to determine the ratios $t_{j,n}/t_{n,j}$.

In all three approaches, note that we  use (\ref{BA2}), rather than (\ref{st}), to 
determine some of the ratios $t_{ij}/t_{ji}$. This is a basic feature of the
algebra of the next two sections.

\subsubsection*{Infinite $v_j$s: the general situation.}

Suppose just $r$  of the $v_1, \ldots , v_n$ equal $- \infty$, $s$ equal $+
\infty$, and the remaining $n-r-s$ are finite and arbitrary. More precisely:
\ba
v_j  & \! \! = \! \!  & \; \; {\rm finite \; , \;   for }\; \; j = 1, \ldots , n-r-s
\comma  \nonumber
\\ v_j  &\! \!   =  \! \!  & \; \;  - \infty  \; , \;   e^{\iota \kappa_j} = q \; , \;   {\rm
for }\; \; j = n-r-s+1,
\ldots , n-s \comma  \\ v_j  & \! \!  = \! \!   & \; \;  + \infty  \; , \;  e^{\iota \kappa_j} =
q^{-1}  \; , \;  {\rm for }\; \; j = n-s+1, \ldots , n \period  \ea

Let us call these three cases type 1 to type 3, respectively. From (\ref{defsjm}) and
(\ref{defq}), if $v_j$ is of type 2 and $v_m$ is not of type 2, then
\be s_{jm} = (q- e^{\iota k_m} ) /q \sep s_{mj} = -q (q- e^{\iota k_m} ) \ee
so  $s_{mj} /s_{jm} = t_{mj} /t_{jm} = -q^2$. Similarly, if $v_j$ is of type 3 and $v_m$ is
not of type 3, then $s_{mj} /s_{jm} = t_{mj} /t_{jm} = -q^{-2}$.

If $v_j$  and $v_m$ are both of type 2, or both of type 3, then $s_{jm} = s_{mj} = 0$ and 
$A(P)$ is not necessarily given by (\ref{Asoln}). Strictly, we should go back to the
original Bethe ansatz equations (\ref{aperm}) and (\ref{acyc}).

However, for similar reasons to those above, it seems that we can without loss of generality
take $A(P)$ to be given by (\ref{Asoln}), so long as we realise that (\ref{st}) no longer
defines $t_{jm}/ t_{mj}$ for  $v_j$  and $v_m$  both of type 2, or both of type 3. Then
(\ref{aperm}) gives the equation (\ref{BA2}). Taking  $v_j$ to be of type 1, 2 or 3,
we obtain
\ba \label{threeqns} e^{i {\L} k_j }  = &  q^{2s-2r}
 \prod_{ m=1  } ^{n-r-s} \left(  - {s_{m,j} / s_{j,m}  }\right) \; \; , \;
 &  \; \;  j = 1,  \ldots , n \! - \! r \! - \! s \; , \nonumber \\
e^{2 \L H} q^{\L}  = &  q^{2n-2r}
 \prod_{ m=n-r-s+1  } ^{n-s} \left(  - {t_{m,j} / t_{j,m}  }\right)  \;  ,
&  j = n \! - \! r \! - \! s \! + \! 1,  \ldots , n \! - \! s \; ,  \nonumber\\
e^{2 \L H} q^{-\L}  = &  q^{2s-2n}
 \prod_{ m=n-s+1  } ^{n} \left(  - {t_{m,j} / t_{j,m}  }\right)  \; \; ,
 &  \; \;  j = n \! - \! s \! + \! 1,  \ldots , n \; ,  \ea
where all three products exclude the value $m = j$.

Taking the product of each of these three equations over the allowed values of $j$, the
$- s_{m,j} / s_{j,m}$, $- t_{m,j} / t_{j,m}$ factors cancel out, leaving
\bd
e^{i {\L} (k_1 + \cdots + k_{n \! - \! r \! - \! s  } )} \eq q^{2 (s \! - \! r)(n \! - \! r
\! -  \! s) }\comma \ed
\be \label{condqH}
e^{2 \L r H} = q^{r(2n-2r-\L)} \sep e^{2 \L s H} = q^{s(\L +2s-2n)} \period \ee
Eliminating $H$ gives
\be
q^{2 r s (\L +r +s - 2 n) } = 1 \comma \ee
so we see that such infinite zeros can only occur when $r$, $s$ or $\L + r +s - 2n$
vanishes, or when $q$ (and $e^{2 H}$) is a root of unity.

Let $\hat{Q}(v)$ be given by (\ref{qdef}), but with the product restricted to the finite
$v_j$:
\be
\hat{Q}(v) \eq \prod_{j=1}^{n \! - \! r \! - \! s} \sinh [(v-v_j)/2] \comma \ee
then, to within factors independent of $v$,
\be
Q(v) \eq e^{ (r-s) v/2} \, \hat{Q}(v) \period \ee
Substituting into the eigenvalue equation (\ref{eig2}), we obtain
\be \label{eig2b}
\Lambda \eq (-1)^n \; \frac{\omega \, \phi (\lambda-v)\, \hat{Q}(v + 2 \lambda) +
\omega^{-1} \, \phi (\lambda+v) \, \hat{Q}(v - 2 \lambda) }{ \hat{Q}(v)} \comma \ee
where
\be \omega = e^{\L H} \, (-q)^{r-s}  \period \ee
Hence
\bd \omega^{2r} = q^{r (2n-2s - \L)} \sep \omega^{2s} = q^{s (\L +2r -2n)} \comma\ed
and we see that if $r$, $s$, $\L + r +s - 2n$ are all non-zero, then $\omega$ must be a root
of unity.

\section{The zero-field model: $H = 0$} 

   The next step was to turn off the field, setting $H=0$. For ${\L}$ odd no
problems appeared: we were able to calculate all the eigenvectors in all the subspaces $n = 0,
\ldots , {\L}$ without difficulty. For
${\L}$ even we encountered two problems, both of which have been discussed previously in the
literature.

\subsubsection{``Beyond the equator'': ${\L}$ even and $n > {\L}/2$. $Q(v)$ has infinite zeros}

In this ``beyond the equator'' case\cite{Pronko99}, (\ref{Q1Q2}) has the simple solution
\be Q_2(v) = Q_1(v)  \period \ee

Certainly (\ref{bqeqn}) permits this solution, and we observe numerically that $\bf B$ has
nullity one, so it is the only solution. Hence
$Q(v)$ is the same for $n$ as for
${\L}-n$. This is quite consistent with (\ref{qdef}) - it merely means that $2n-{\L}$ of the zeros
have gone off to infinity. This is even easier to see in (\ref{Qexp}): the first  $n-{\L}/2$
coefficients
$q_j$ vanish, as do the last $n-{\L}/2$ coefficients. The degree of the Laurent polynomial is
reduced from $n$ to
${\L}-n$, which is  still consistent with  (\ref{bqeqn}).

From (\ref{eik}) and (\ref{defq}), this means that  $n-{\L}/2$ of the wave numbers $k_1, 
\ldots, k_n$ are given by $e^{\iota k_j } = q$, and another  $n-{\L}/2$ by $e^{\iota k_j } = 1/q$.

Faddeev and Takhtajan \cite{FaddTak84}  state that ``Bethe's
vector vanishes'' for $n > {\L}/2$.  For $n = 1 +  {\L}/2$ this is certainly not true of the vector $g$
as given by (\ref{gexp}) and (\ref{Asoln}). We have observed numerically that it is non-zero
and that it is indeed the eigenvector of $T$.

For $n > 1 +  {\L}/2$ there is a problem, but it can be overcome, using the working at the
end of the last section, taking $r, s$ therein to be $n - \L/2$, so that (\ref{condqH}) is
satisfied for all $\lambda, q$ . When $v_j = v_m = \pm \infty$, then $s_{m,j} = s_{j,m} =
0$, so (\ref{st}) tells us nothing about $t_{m,j}, t_{j,m}$: instead they
should be chosen to satisfy the last two equations of (\ref{threeqns}). There will be many ways
to do this, corresponding to the fact that $A(P)$ enters (\ref{gexp}) only via its sum over all
ways of permuting $v_{\L- n+1}, \ldots , v_{\L/2}$ and  $v_{\L/2 +1}, \ldots , v_{n}$. One
simple way is to take $t_{m,j} = - t_{j,m}$. The $\L -n$ finite zeros $v_j$ are given by the
first of the equations (\ref{threeqns}), which is the same as (\ref{BA2}) when $n$ is replaced
by $\L - n$, i.e. the Bethe equations for the {\em right} side of the equator. The eigenvalue
equations are the same for both $n$ and $\L - n$.

The eigenvector equations are different, since we must include all $n$ zeros in
the product in (\ref{Asoln}). The resulting coefficients
$A(p_1, \ldots ,p_n)$  are finite and non-zero. The equations (\ref{aperm})
and (\ref{acyc})  are  satisfied, so the vector
$g$ with elements (\ref{gexp}) must be an eigenvector if it is non-zero. It is non-zero in our
numerical experiments, and it is the eigenvector corresponding to the eigenvalue $\Lambda$.

Thus despite the assertions that have been made in the past, the Bethe ansatz can be used to
construct the eigenvector $g$ for  $n > {\L}/2$, even when ${\L}$ is even. It is a 
furphy that Bethe's
ansatz does not work on the wrong side of the equator.

\subsubsection{${\L}$ even and $2 \leq n \leq {\L} -2$: a single bound pair} 

The other problem that we encountered first occurs for ${\L}= 4$ and $n=2$, then for even ${\L}$ and
$2 \leq n \leq {\L}-2$. It is referred to by Bethe himself \cite[after eqn.
23]{Bethe31} and has been considered by others since 
\cite{Sidd98}, \cite{NohLeeKim00},\cite[eqn.3.2.23b]{Batchelor87}.
 For some eigenvalues with
momentum $\pm 1$, i.e  $k_1 +
\cdots + k_n = 0$ or
$\pi$, we found that $Q(v)$ had a pair of zeros $v_1, v_2$ such that $v_1 = \lambda$, $v_2 =
-\lambda$. From (\ref{eik}) this implies that
\be \label{boundpair}
e^{ \iota k_1} =  e^{ -\iota k_2}  =  0 \period \ee
More strongly, it was always true for such pairs that
\be e^{\iota(k_1+k_2)} = -1    \comma \ee
so from (\ref{sdef2}),
\be s_{12} = 2 \Delta e^{-\iota k_1}  = \infty \ee
while $s_{21}$ vanishes. 

Since $s_{12}, s_{21}$ do not both vanish, we can take as usual $t_{12} = s_{12}$, 
$t_{21} = s_{21}$.

In fact $t_{21}$ vanishes strongly as this situation is approached (say by turning off the field
$H$): from (\ref{BA2}),
\be t_{21} = C_{21} \, e^{\iota({\L}-1) k_1 } \comma \ee
where $C_{21}$ has a finite non-zero value given by (\ref{BA2}).

There is no problem solving Bethe's equations for the other $k_3, \ldots , k_n$ and for $C_{21}$.
In principle one can then substitute these expressions into (\ref{Asoln}) and (\ref{gexp})
and extract the terms and elements that grow most rapidly as $ e^{ i k_1}$ vanishes.
Choosing $C$ as stated after (\ref{st}), each term in the summand of (\ref{gexp}) will
give a  finite contribution to the eigenvector $g$. Some will be zero, but our numerical
experiments indicate that the total vector $g$ is not  zero, and is in fact the correct
eigenvector. 

\subsubsection*{Numerical results} 

For our numerical experiments, we simply assigned  $ e^{
\iota k_1}$ the numerically small but non-zero value $10^{-14} \iota$ (the $\iota$ being
necessary for
$k_1$,
$k_2$ to be complex conjugates), calculated the $s_{ij}$  other than
$s_{21}$ from (\ref{sdef2}), then calculated $s_{21}$ from (\ref{BA2}), substituted the
results into (\ref{Asoln}) and (\ref{gexp}) and normalized the vector $g$. We obtained the
correct eigenvector to approximately 14-digit accuracy.

For ${\L}=4$ there was just one such eigenvalue $\Lambda$, in the $n = 2$ central block. For ${\L}=6$
there was one in the $n=2$ block, two in the $n=3$ block, and one in the $n=4$ block. For ${\L}=8$
there were $1, 2, 5, 2, 1$ in the n = $2,3,4,5,6$ blocks, respectively. This suggests
(tentatively) that the Catalan numbers may count such eigenvalues. The momenta were $-1$, except
for a single eigenvalue with momentum $+1$ in each block with $3 \leq n \leq {\L}-3$.

When $ {\L}/2 +1 < n < {\L}-1$, there are eigenvalues where the two problems we have just discussed occur
together, i.e. more than one of the $e^{i k_j}$ are equal to $q$, more than one to $1/q$, and two
of the others are $\pm i \infty $. The above procedures were built into our computer program and
worked perfectly, giving the correct non-zero eigenvector of $T$.

For $\L$ even, we also calculated the matrix $\Phi_n(v )$ of equation (96) of \cite{Baxter73a}.
We found that its rows were, as expected,  linearly independent,  so we were able to use
(\ref{cnstrctQ}) above  to calculate 
${\tilde Q}(v)$, and did indeed find that this matrix was diagonalized by the same matrix
$P$ that diagonalizes
$T(v)$, and that each eigenvalue was  the function $Q(v)$ discussed above (more precisely,
the eigenvalue  was  $Q(v)/Q(v_0)$ ).



The main lesson from this and the previous section is that Bethe's equations should be viewed as a
set of coupled non-linear equations for the
$e^{\iota k_j}, s_{ij}$ and the ratios $t_{ij} : t_{ji}$. We do not necessarily proceed by solving
(\ref{sdef2}) and (\ref{st}) for the ratios  $t_{ij} : t_{ji}$. For some $i$ and $j$ it may be
appropriate to obtain this ratio from (\ref{BA2}).

\section{$H = 0$ and $q$ a root of unity: exact complete strings}

Now we come to the case discussed by Deguchi, Fabricius and McCoy in \cite{DegFabMcCoy01} -
\cite{FabMcCoy01c}, where $\lambda $ is a rational fraction of $\iota \pi$, i.e.
there exist  integers $\nu, M$ (with no common factors) such that 
\be  \label{lamval}   \lambda = i \nu \pi /M \sep q = - e^{i \nu \pi /M} \comma \ee
using (\ref{defq}).

Then 
\be\label{qpwr}
q^{2 M} = 1 \comma \ee
and there is no smaller integer power of $q^2$ that equals one.

For the moment we allow $H$ to be arbitrary: we shall show that the further restriction
(\ref{restrH}) is necessary for a  string (more precisely, a single string) to
occur, and from then on take $H$ to be zero.

The set of Bethe zeros $v_1, \ldots , v_n$ may now contain one or more  ``complete
strings'', in which $M$ of them, say  $v_1, \ldots , v_{M}$, are related by
\be \label{string}
v_{j} = v_1 + 2 (j-1) \lambda \; \; \; {\rm for } \; \; \; j=1, \ldots , M
\period \ee 
This implies that 
$ v_{j+1} = v_j + 2 \lambda $ and $v_1 = v_M  + 2 \lambda  - 2 \iota \nu \pi $, so from
(\ref{defs2}),
\be \label{sijzero}
s_{12}  = s_{23} = \cdots = s_{M - 1,M} = s_{M,1} =  0 \period \ee
We see that some of the $s_{ij}$ vanish, so we have to be careful with Bethe's equations.

The set $\{ v_1 ,\ldots , v_M \}$ may of course contain more than one complete string.
For simplicity, from now on we shall restrict our attention to the case when there is only one
string, but we fully expect our methods and comments to be applicable to the general case.

For $j=1, \ldots , M$, the equation (\ref{BA}) becomes $0 = 0$, which is true, but not helpful.
For $j  = M+1, \ldots ,n$, both sides of the equation are non-zero. Using the form  (\ref{defs2})
of $s_{ij}$, we obtain
\be
s_{1,j} s_{2,j} \cdots s_{M,j} =  (-1)^M s_{j,1} s_{j,2} \cdots s_{j,M} \comma \ee
so for $j> M$, equation  (\ref{BA}) simplifies to
\be \label{BAred2}
e^{i \L k_j} \eq (-1)^{n-1-M} \prod_{m=M+1, m \neq j}^n s_{m,j}/s_{j,m} \period \ee
Taking the product of these $n-M$ equations, the $s_{j,m}, s_{m,j}$ factors cancel, leaving
\be
e^{i\L (k_{M+1} + \cdots + k_n)} \eq 1 \period \ee
From (\ref{sumks}) it follows that
\be \label{restrsum}
e^{i\L (k_{1} + \cdots + k_M)} \eq 1 \period \ee

Now we note from (\ref{string}) and (\ref{eik}) that, for $j = 1, \ldots , M$,
\be \label{eik2}
e^{ik_j} \eq q e^{2H} \frac{1+q^{2j-3} z_1 }{1+q^{2j-1} z_1 } \comma \ee
where $z_1 = \exp (v_1)$ and more generally
\be
z_j \eq e^{v_j} \eq q^{2j-2} z_1 \period \ee

Substituting this into (\ref{restrsum}) and using (\ref{qpwr}), we obtain
\be \label{condn2}
q^{\L M} e^{2 \L  M H} = 1 \period \ee

This is an extra condition on $q$ and $H$ that must be satisfied for a complete string to
occur.\footnote{At least for just one string to occur, but the same condition appears to be
necessary for any number of strings.} From this and (\ref{qpwr}) we see that
\be \label{restrH}
e^{4 \L  M H } = 1 \period \ee
An obvious and interesting solution of this equation is
\be \label{Hzero}
H = 0 \comma \ee
and from now on in this section we shall take $H=0$, but we do note that there are other (pure
imaginary) values of $H$ for which complete strings may occur.

If ${\L}$ is even, (\ref{condn2}) is implied by (\ref{qpwr}). If ${\L}$ is odd, the two
equations together imply that $q^M = 1$, but this is consistent with $q^2$ being a primitive
$M$th root of unity only if $M$ is odd. Thus we have two possibilities:
\ba \label{constr}
q^{2 M} \eq 1 & \sep &  {\L} \; \; {\rm even} \comma \nonumber \\
 q^{M} \eq 1 & \sep &  {\L} \; {\rm and} \; M \; {\rm both \; \; odd} \period \ea

\subsubsection*{Apparent difficulties: (1) calculating $g$ } 

There are two problems that appear when a complete string occurs: the Bethe equations do not
have a unique solution, and if we use the obvious solution $t_{ij} = s_{ij}$ of (\ref{st}) in
(\ref{Asoln}), then every coefficient $A(P)$ vanishes, so the eigenvector $g$, given by
(\ref{gexp}) also vanishes.\footnote{The only way the rhs of (\ref{Asoln})
could be non-zero would be for it to contain the factors $s_{21}, s_{32},
\ldots, s_{M ,M-1}, s_{1,M}$.  But this cannot happen as there is no inverse
permutation $P'$ such that $p'_2 > p'_1, p'_3 > p'_2, \ldots , p'_1 > p'_M$: 
the inequalities are inconsistent. At least one of them must fail, which is
the reason for the renormalization of $A(P)$ proposed below.} 

Let us dispose of the second difficulty first, since it is fairly straightforward. For the
algebraic Bethe ansatz, it is considered by Fabricius and McCoy in the remarks after their
equation (1.36) of \cite{FabMcCoy01c}.

We can still use the
ansatz (\ref{Asoln}) for the coefficients $A(P)$ and attempt to satisfy the modified Bethe's
equations (\ref{st}), (\ref{BA2}). Taking
$t_{ij} = s_{ij}$ for all $i,j$, we note from (\ref{sijzero}) that 
\bd t_{12} = t_{23} =  \cdots  = t_{M,1}  = 0 \period \ed 

 For $j = M+1,
\ldots ,n$, (\ref{BA2}) becomes  the above reduced equations (\ref{BAred2}), which can be
viewed as fixing $v_{M+1}, \ldots ,v_n$.

Let $v_1$ be assigned arbitrarily. Then $v_2, \ldots , v_M$ and $e^{i k_1}, \ldots , e^{ik_M}$
are given by (\ref{string}) and (\ref{eik2}).

We still have to satisfy (\ref{BA2}) for $j=1, \ldots ,M$. We can do this by using it to determine
the ratios $t_{j-1,j}/t_{j,j+1}$, for $j = 1, \ldots ,m$  (with
$t_{0,1}  = t_{m,m+1} = t_{m,1}$). We could of course have originally formulated
Bethe's equations (which are just a set of algebraic equations and may well have solutions at
zero or infinity) in terms of these non-zero, finite ratios.

We also take $C$ in (\ref{Asoln}) to be $t_{12}$. Then some of the coefficients $A(P)$ will
depend on  $t_{12},t_{23}, \ldots , t_{M,1}$ only via their ratios, which are given by
(\ref{BA2}). The remaining coefficents $A(P)$ will vanish. 

Thus all the coefficients are finite, some are non-zero, and we may hope that the eigenvector
$g$, given by (\ref{gexp}), will befinite and  non-zero. In our numerical experiments this is
what we have found.

If there are $\gamma$ complete strings, then $C$ should be the product of $\gamma$
factors $t_{ij}$, being one of the vanishing $t_{ij}$ from each string.

\subsubsection*{Apparent difficulties: (2) non-uniqueness of the  eigenvector } 

We said above ``let $v_1$ be assigned arbitrarily''. Why is this allowed? Should not its
value be determined? This is the problem that concerned Deguchi, Fabricius and
McCoy.\cite{DegFabMcCoy01} - \cite{FabMcCoy01c}

The answer is that for any value of $v_1$ the above procedure satisfies (\ref{Asoln}) -
(\ref{BA2})  and therefore the Bethe ansatz equations (\ref{gexp})  - (\ref{sumks}).
Provided $g$ is not zero (and our numerical calculations indicate that it is not), then it
must be an eigenvector of the transfer matrix $T(v)$. We are free to choose $v_1$ as we
wish. There is no {\it a priory} need to ``complete'' Bethe's equations.\cite{FabMcCoy01b}

Let us look more closely at what is happening.

 From
(\ref{string}) and (\ref{qdef}), the $M$ zeros $v_1, \ldots ,v_M$  contribute to $Q(v)$ 
a factor
\be  \prod_{j=1}^M \sinh [(v-v_j)/2]  \propto \sinh [M(v - v_1)/2]  \period \ee
This factor cancels out of 
(\ref{eig2}), except only for a constant $(-1)^{\nu}$.

It follows at once that if $Q(v)$ satisfies (\ref{eig2}), and hence the Bethe equations, {\em then so
will any other function $Q(v)$ with a different value of $v_1$.}

There is nothing remarkable about this. It means that the matrix $\bf B$  has rank
less than $n$, so there is more than one solution $q_0, \ldots, q_n$ of (\ref{bqeqn}). This in turn
is a signal that the eigenvalue $\Lambda(v)$ is degenerate, for {\em all} $v$.

Of course, changing $v_1$ will change $Q(v)$, so the matrix ${\tilde Q} (v)$ will not be degenerate.
If we construct it explicitly as above, then diagonalizing ${\tilde Q} (v)$ rather than $T(v)$
will resolve the degeneracies of $T(v)$. Further, these will give the eigenvalues and eigenvectors
obtained by taking the limit as $\lambda$ approaches the value (\ref{lamval}). This is what
Fabricius and McCoy have done, and the results are interesting. 

What we are concerned with here is showing that there is nothing in their work that indicates that
the Bethe ansatz is incomplete, in the usual sense of not giving all the eigenvectors or ``states''.
The fact that $Q(v)$ is not uniquely defined by (\ref{gexp}) - (\ref{sumks}) is precisely the reason
why one can use these equations to obtain a complete basis of the eigenspace of the eigenvalue
$\Lambda (v)$.

{\bf A single string containing all the $v_1, \ldots , v_n$.} ~~\\

For simplicity we further restrict our attention to the case when {\em all} the $v_1, \ldots
,v_n$ lie in the string, i.e.
\be M = n \period \ee
However, we expect the substance of our remarks to generalize to $n >M$, and indeed to the case
when $v_1 ,\ldots ,v_n$ contain more than one string.

The eigenvalue $\Lambda$ is given
immediately by (\ref{eig2}):
\be \label{eig2a}
\Lambda \eq q^n [ \phi (\lambda-v) + \phi (\lambda+v) ] \eq q^n (a^{\L}+b^{\L})
\period\ee
To within a sign, it is the eigenvalue for the ``vacuum'' state, when all spins are up.
We shall now use the original Bethe ansatz equations (\ref{gexp})  - (\ref{sumks}) to obtain
explicit expressions for all the eigenvectors
$g$ corresponding to this eigenvalue, for a given value of the number $n$ of down arrows. We
shall {\em not} use ``Bethe's equations'' (\ref{Asoln}) - (\ref{BA2}).

From (\ref{string}) and (\ref{defs2}),
\be \label{szero}
s_{12} =  s_{23} = \cdots = s_{n-1,n} = s_{n,1} = 0 \period \ee

The coefficients $A(1,2,\ldots ,n), A(2,3,\ldots,n,1), \ldots , A(n,1,2,\ldots ,n-1) $
enter the equations (\ref{aperm}) only with  multiplying factors $s_{ij}$ that belong to the
set (\ref{szero}). This means that these coefficients do not enter at all. It appears that all
other coefficients do enter with non-zero multiplying factors, and the equations ensure that they
vanish.\footnote{I have only verified this for $n = 2,\ldots, 9$, but this strongly suggests
that it is correct.}

Choosing $A(1,2,\ldots ,n) = 1$, from (\ref{acyc}) it follows that
\be \label{Aspeccyc}
A(j,j+1,\ldots , n,1,\ldots ,j-1) \eq e^{i {\L}(k_j+k_{j+1} + \cdots +
k_n)}  \period \ee

 From (\ref{gexp}) it follows that
\bd  g(\x_1,\ldots ,\x_n) \eq
\sum_{j=1}^n \exp \{ i [k_1 \x_{n+2-j} + k_2 \x_{n+3-j} + \cdots \ed
\be \label{gexp2} \cdots  + k_{j-1} \x_n +
k_j (\x_1 \! + \!  {\L}) + k_{j+1} (\x_2 \!  + \!  {\L}) + \cdots + k_{n} (\x_{n+1-j} \!  + \!  {\L})] \}
\period
\ee
The cyclic property $ g(\x_1,\ldots ,\x_n) =  g(\x_2,\ldots
,\x_n,\x_1+{\L})$ is  manifested by (\ref{gexp2}).

All of the Bethe ansatz equations (\ref{gexp}) - (\ref{eigval}) are now satisfied.
We still have the parameter $v_1$, or equivalently $z_1$,  at our disposal.

Substituting (\ref{eik2}) into (\ref{gexp2}) and dividing by a common
normalisation factor $(1-(-z_1/q)^n)/ n (1+z_1/q)^{\L} $, we obtain
\be \label{gexp3}
g(z_1 | \x_1,\ldots ,\x_n) \eq n^{-1} \sum_{j=1}^n q^{\x_1 + \cdots \x_n
+{\L}(1-j)} \prod_{r=1}^n (1+ q^{2j+2r-3} \, z_1)^{\x_{r+1}-\x_r-1} \comma \ee
taking $\x_{n+1} = \x_1+{\L}$ and exhibiting the dependence of $g$ on $z_1$.

From (\ref{constr}), $q^{\L n } = 1$, which means that the summand in (\ref{gexp3}) is
unchanged by replacing $j$ by
$j+n$. Replacing $j, z_1$ by $j-1, q^2 z_1$, we observe that
\be \label{greln}
g(x^2 z_1 | \x_1,\ldots ,\x_n) \eq q^{\L} \, g(z_1 | \x_1,\ldots ,\x_n) \period \ee

Because of
 the restrictions (\ref{range}), the RHS of (\ref{gexp3}) is a polynomial in $z_1$ of
degree ${\L}-n$. Hence we can expand the vector $g$ and its elements in powers of $z_1$:
\bd g(z_1) \eq \sum_{k=0}^{{\L}-n}  \, z_1^k \, c_k \comma \ed
\be \label{gpoly} 
g(z_1 |\x_1,\ldots
,\x_n)   \eq \sum_{k=0}^{{\L}-n}z_1^k \,  c_k(\x_1,\ldots
,\x_n)   \comma \ee
$c_k$ being a vector with elements $c_k(\x_1,\ldots,\x_n) $.

Substituting the expression (\ref{gpoly}) into (\ref{greln}), we find that
\be \label{restrk} c_k = 0 \; \; \; {\rm unless} \; \; 
  q^{2 k}  = q^{\L}   \period \ee
This means that most of the $c_k$ vanish. Define an integer $\alpha$ by
\ba
\alpha = {\L}/2 \; ,   & {\rm mod} & n \; \; \; {\rm if } \; {\L}  \; {\rm is \; \;   even }
\nonumber \\
\alpha = ({\L}-n)/2 \; ,  & {\rm mod} &  n \; \; \; {\rm if } \;  {\L}, n  \; \, {\rm are
\;  odd }
\comma \ea
so that in either case $0 \leq \alpha  < n$. Then $c_k$ is non-zero only when $k = \alpha,
\alpha + n, \alpha + 2 n, \ldots $. It follows that there are at most
\be \label{degeneracy}
{\cal N} = \left[ \frac{{\L} - \alpha }{n}\right] \ee
non-zero vectors $c_k$ in the expansion (\ref{gpoly}). Here $[x] $ denotes
the integer part of $x$.

We can write down an explicit, if unwieldy,  expression for the elements of $c_k$
by performing a binomial expansion on each product in (\ref{gexp3}):
\be  \label{cks}
c_k(\x_1,\ldots
,\x_n)   =   q^{\x_1 + \cdots + \x_n} \, \sum_{m_1 , \ldots ,m_n} 
 \prod_{r=1}^n q^{(2 r-1) m_r} \left( \begin{array} {c} \! \! \x_{r+1}-\x_r - \! 1 \! \!\\ m_r
\end{array} \right) \comma   \ee 
where $k$ must satisfy the restriction (\ref{restrk}) and the summation is over all integers
$m_1 ,\ldots ,m_n$ such that
$m_1 + \cdots + m_n = k $ and 
\bd
0 \leq m_r \leq \x_{r+1}-\x_r -1  \; \; , \; \;{\rm for } \; r = 1, \ldots , n \period \ed 

\subsubsection{Numerical tests} 

It is not obvious whether these vector are in fact linearly independent. The author
knows of no reason to suppose they are not, but as a check we have numerically
calculated the vectors for $ n = 2$ and 3. We can distinguish three cases:

i)   $n=2$ ,  $\lambda = i \pi/2$,  $q = -i$, $q^n = -1$,

ii)  $n=3$ ,  $\lambda = i \pi/3$,  $q = e^{-2 \pi i/3}$, $q^n = 1$,

iii)  $n=3$ ,  $\lambda = 2i \pi/3$,  $q = e^{- \pi i/3}$, $q^n = -1$.

For all these cases $q^{2n} = 1$, but only for the second is $q^n=1$. Thus ${\L}$ must be even for
cases (i) and (iii), but may be either even or odd for case (ii).

We present the  results in Table 1, for ${\L} = 2, \ldots ,
16$. In every case we calculated  the vector $g$ with elements (\ref{gexp3}) for 12 randomly
chosen values of
$z_1$ and then determined (to 15 digit precision) the rank $\tilde{r}$ of the matrix with these 12
column  vectors. We then numerically verified that each was an eigenvector of the six-vertex
model  transfer matrix $T$, with eigenvalue  (\ref{eig2a}). Finally, we calculated the column
nullity $\tilde{n}$ of $T - \Lambda I$ in the subspace with $n$ down arrows. This is the
degeneracy of
$\Lambda$. In every case we found $\tilde{r} = \tilde{n} = {\cal N}$, where $\cal N$ is given
(\ref{degeneracy}). Thus at least for these cases $
\cal V$ is indeed of dimension (\ref{degeneracy}) and contains all eigenstates.

\begin{table} ~~~~~~~~~~~~~~~~~~~~~~~
{\renewcommand{\arraystretch}{1.3} 
{\flushleft{\hspace{0.7 cm}
\begin{tabular}{|c|ccccccccccccccc|}
\hline{\L} &  2 & 3 & 4 & 5 & 6 & 7 & 8 & 9 & 10 & 11 & 12 & 13 & 14 & 15 & 16\\ 
\hline
case (i)    & 0  &    & 2  &   & 2 &   & 4 &   & 4 && 6 && 6 && 8 \\ 
case (ii)   & 0  & 1  &  0 & 1 & 2 & 1 & 2 & 3 & 2 & 3 & 4& 3 & 4 & 5 & 4   \\
case(iii)   & 0  &    &  0 &   & 2 &   & 2 &   & 2 && 4 && 4 && 4\\
 \hline
\end{tabular} 
}}
\caption{{\protect{\footnotesize The dimensions of the space $\cal V$ for $n=2$ and
$n=3$, as  calculated numerically (a value of zero implies that their are no eigenvectors and
that (\ref{eig2}) is not an eigenvalue). In every case the result agrees with (\ref{degeneracy})
and with the calculated nullity of $T -
\Lambda I$, implying that the vectors (\ref{cks}) are indeed linearly independent, and that
$\cal V$ is the complete eigenspace.}}} }
\end{table}

We can compare these results with those of Fabricius and McCoy
\cite{FabMcCoy01a}. Their $\gamma$ is related to our $\lambda$ by $\lambda = i \gamma$, so 
$\Delta = - \cos \gamma$ and $q = - e^{i \gamma}$.
 From Table 2 of their paper, when $\gamma = \pi/2 $, $\Delta = 0$, $n=2$ and ${\L}=16$, there
are eight single strings in the $S^z = {\L}/2 - n = 6$ sub-space, all corresponding to the same
eigenvalue. This agrees with the above derivation  : $\alpha
= 0 
$ , so
${\cal N} = 8$ and the summand in (\ref{gpoly}) is non-zero only when
$k$ takes the eight values $0,2,4,6,8,10,12,14$.

Also, from Table 7 of the same paper, when
$\gamma = \pi /3$, 
$\Delta = -1/2$, $n = 3$ and  ${\L}=16$, there are four equal-eigenvalue single strings
in the 
$S^z = {\L}/2 - n = 5$ sub-space. This also agrees with the above, and with case (ii) in the 
table: $\alpha =  2$,  ${\cal N} = 4$,  and  $ k = 2, 5, 8, 11$. 

Some of these results have also been obtained by Braak and Andrei \cite{Braak01}, who refer to the
freedom in the choice of the string centres as ``transparent excitations''. Their table 1 is line
2 of our Table 1 above.

\subsubsection{Conclusions} 

 Let
$\cal V$ be the space that spanned by the non-zero vectors $c_k$. Then as $z_1$ varies,  the
eigenvector
$g(z_1)$ traces a curve within this space. Each vector $c_k$ can be written as a sum of eigenvectors
$g(z_1)$, so is itself an eigenvector. $\cal V$ is an eigenspace corresponding to the
eigenvalue (\ref{eig2a}).

 It appears that the vectors $c_k$ are linearly independent and span the full eigenspace (within
the sub-space $S^z = {\L}/2 - n$ of $n$ down arrows). If so, then $\cal V$ is 
of dimension $\cal N$. The eigenvalue has degeneracy  $\cal N$, and we have used the Bethe
ansatz to construct the full eigenspace.

\subsubsection{The case when the string parameters  $v_1, \ldots , v_m$ are infinite}

If $q$ satisfies both the 
restrictions (\ref{constr}) and 
\be  \label{restrq}
q^{\L} = 1 \comma \ee
then $k=0$ and
$k= {\L}-n$  in (\ref{gpoly}) both correspond to non-zero vectors $c_k$. These $c_k$ are the
values of $g$ when $z_1 = 0$ and $z_1 = \infty$, i.e. when $v_1, \ldots, v_m$ = $ - \infty$ and
 $v_1, \ldots, v_m$ = $+\infty$. We have a string of infinite $v_1, \ldots , v_m$.

From (\ref{gexp3}),
we readily find that
\ba
c_0(\x_1,\ldots ,\x_m)  & \eq  &  q^{ \x_1 + \x_2 + \cdots + \x_m} \comma \nonumber \\
c_{{\L}-m} (\x_1,\ldots ,\x_m) & \eq  & q^{m-{\L}} \, q^{- \x_1 - \x_2 - \cdots -
\x_m} 
\ea
These are particular cases of the special vectors reported in equations (16) - (22) of
\cite{Baxter73a},  in \cite{Jones74}, and in the equation (\ref{ddphi}) above.

\subsection*{Multiple complete strings: the function $Q(v)$}


We also calculated the null space of the matrix $\bf B$ in (\ref{BAalt2}), algebraically using
Mathematica. We fixed the values of $\lambda$ and $q$  as in cases (i), (ii), (iii). We then allowed
$n$ (the number of down arrows) in (\ref{eig2a}) and (\ref{BAalt1}) - (\ref{BAalt2}) to take all
values from 0 to ${\L}$,  for ${\L} = 2,3,\ldots ,9$. Since $\Lambda (v)$ is given by 
(\ref{eig2a}), we could  immediately calculate $t_0,
\ldots ,t_{\L}$ and form $\bf B$. We did indeed find that the column nullity of
$B$ was sometimes greater than one. For cases (ii) and (iii), where $-3i \lambda$ is an integer, we
found that the nullity was zero unless $n$ was a multiple of 3, meaning that $\Lambda (v)$ was not
an eigenvalue . When
$n$ was a multiple of 3 the nullity  was $(n+3)/3$ and (\ref{bqeqn}) was satisfied provided only
that $q_j$ was zero when
$j$ is not a multiply of 3. Thus 
\bd 
e^{nv/2} Q(v) = {\rm arbitrary \; polynomial \; in \;} z^3 \; {\rm of \; degree \;} n/3 \comma \ed
where $z = e^v$. Whatever the choices of the coefficients of this polynomial, it can be factored
into $n/3$ polynomials of degree 1 in $z^3$, each of which is a complete string.

So $Q(v)$ factors into a product of $n/3$ complete strings of length 3. The `` centre'' 
(the average value of the three $v_j$s in the string) of each string is undetermined. Any such
function $Q(v)$ satisfies (\ref{bqeqn}).

We found corresponding behaviour for case (i), when $\lambda = i \pi/2$: for $n$ even, $Q(v)$
factors into a product of $n/2$ undetermined complete strings of length 2. There no solutions for
$n$ odd.

We have not attempted to generalize the derivation of this section of the eigenvector $g$ to such
multiple strings, but presume that it can be done, and that one would find the remarkable binomial 
pattern of degeneracies reported by Fabricius and McCoy in their tables 2 and 7 for `maximum $S^z =
8$'.\cite{FabMcCoy01a}

In the $2n =  \L$ sub-space one can impose the ``sum-rule'' constraint (\ref{sixvconstr}) on $v_1 +
\cdots + v_n$. (For non-degenerate eigenvalues, with no strings, this will automatically be
satisfied. If there are strings, it is not necessary, but may be convenient, and will still give a
complete set of eigenvectors.)
If $v_1, \ldots, v_n$ contain only  one complete string, then this condition can be used to fix
its centre. For $n = 2$ or 3, and ${\L} = 2 n$, this gives $z_1^{n} = 
(-1)^r \exp (-n^2  \lambda) $. This $r$ is 0 for states symmetric under arrow reversal, 1 for
antisymmetric states. We have verified numerically that the resulting eigenvectors (\ref{gexp3}) do
indeed have this (anti-)symmetry.

\section{Some of the $v_1, \ldots , v_n$ equal}

Suppose that all the $s_{i,j}$ are non-zero. Then the $A(p_1, \ldots , p_n)$ are  given by
(\ref{Asoln}). Substitute this into (\ref{gexp}) and for the moment regard $v_1, \ldots , v_n$
as arbitrary parameters and $x_1, \ldots , x_n$ as fixed. The result is an entire anti-symmetric
periodic function of 
 $v_1, \ldots , v_n$. It must therefore contain the factor
\be \prod_{1 \leq i < j \leq n} \sinh [(v_i - v_j)/2] \period \ee
In principle one can divide this factor out (it is the same for all $x_1, \ldots , x_n$, so is
just a normalization factor for the vector $g$). The result is a symmetric function of $v_1,
\ldots , v_n$. For instance, if $n = 2$ and $N=4$,  as  in (\ref{Qexp}) we can write $Q(v)$ as a
Laurent polynomial  in $z = e^v$:
\be
Q(v) = z^{-1} (d z^2 + e z + f)  \period \ee


We can then follow this procedure so as to write all the $g(x_1,x_2)$ as multinomials in
$d,e, f$. Define
\bd
\Delta' \eq 2 \Delta\eq q + q^{-1} \comma \ed
\bd
\xi_1 = q^{-1} d -e + q f \sep \xi_2 = q d -e + q^{-1} f \comma \ed
\bd \xi_3 = (d + f) \Delta' - 2 e \comma \ed
\bd \xi_4 = -4 e (d+f) + (d^2+6 d f + f^2+e^2) \Delta' - 
d f {\Delta '}^3 \comma
\ed
\bd \xi_5 = 2 e (e^2-3 d^2-10 d f - 3 f^2)+(d+f) {\Delta '} (d^2+14 d f +
f^2 +3 e^2) - \ed
\bd
e {\Delta '}^2 (2 d f + e^2) -3 d f (d + f ) {\Delta '}^3 + 
d e f  {\Delta '}^4 \comma \ed
then we find that we can normalize the vector $g$ so that
\bd g(1,2) = \xi_1^2 \xi_3  \sep g(1,3) = \xi_1 \xi_4 \comma \ed
\be \label{gvals} g(1,4) =  \xi_5  \sep g(2,3) = \xi_1 \xi_2 \xi_3 \comma \ee
\bd g(2,4) =  \xi_2 \xi_4  \sep g(3,4) = \xi_2^2 \xi_3 \period \ed

Each element is a multinomial of degree 3 in the coefficients $d, e, f$. Such expressions remove
the difficulty that occurs when two or more of the
$v_j$ become equal: in that case the anti-symmetric factor mentioned above vanishes, as does the
right-hand side of  (\ref{gexp}), but the above expressions do not. 

This procedure is equivalent to taking the limit of the ratios of the eigenvector elements $g(x_1,
\ldots ,x_n)$ in (\ref{gexp}) as the $v_j$ approach one another. It also has the conceptual
advantage of making it clear that one does not necessarily have to calculate the zeros $v_1,
\ldots, v_n$ of
$Q(v)$.  Instead one can imagine solving the bilinear equations given in (\ref{BAalt1}) -
(\ref{BAalt2})  for the coefficients of the Laurent polynomial expansions in $e^{v}$ of
$\Lambda(v)$ and $Q(v)$, then using the results in equations such as the above.

Of course, in practice we do not have useful explicit results for the generalizations of
(\ref{gvals}) to arbitrary $n$ and ${\L}$, so for anything other than small $n, {\L}$ one is forced to
use (\ref{gexp}) and (\ref{Asoln}) directly. However, from this point of view there is nothing
remarkable about two of the $v_j$ coinciding: it merely means that $Q(v)$ has a repeated zero.
If all one wants is the eigenvalue $\Lambda (v)$, then the alternate form 
form (\ref{BAalt1}) - (\ref{BAalt2}) of Bethe's equations can be used directly as written.

\section{Extension to the eight-vertex model}

The zero-field eight-vertex model was solved in 1971 by the author\cite{Baxter71,Baxter72} by
extending the functional relation (\ref{fnlreln}) (with $H = 0$) to the eight-vertex model,
provided that
\be 
\L \eq {\rm \; even} \period \ee

This restriction applies throughout this section.

 This functional relation method gives the eigenvalues
$\Lambda(v)$, but not the eigenvectors. In 1972, while at Stony Brook, the author derived
equations for the eigenvectors of the eight-vertex model in a sequence of three
papers\cite{Baxter73a,Baxter73b,Baxter73c}. The basic technique was to convert the eight-vertex
model to an ice-type solid-on-solid model, and then to solve this by an appropriately 
generalized Bethe ansatz.

Here we use the notation of \cite{Baxter73a,Baxter73b,Baxter73c} and prefix the equations by I,
II, III, according to in which of the three papers it appears. Papers II and III consider  the case
when the parameter
$\eta$ satisfies the ``root of unity'' condition (I.9), (II.6.8) or (III.1.9), i.e.
\be \label{rootofunity}
{\Lc } \eta = 2 m_1 K + i \, m_2 K' \comma \ee
where $\Lc, m_1, m_2$ are integers. This is analogous to (\ref{lamval}).
 This restriction
is {\em not} needed in section 6 of I, because  the condition (8) therein is sufficient to ensure
the required cyclic boundary condition from column $N$ to column 1.

Papers II and III further  consider the case when there are integers $\tilde{n}, \nu'$ such that
\be \label{restrnt}
\L - 2 \tilde{n} = \Lc  \nu' \period \ee

Modified elliptic theta functions are introduced:
\ba \label{HvHJb}
H(u)  & = & H_{Jb}(u) \,\exp [i \pi m_2 (u-K)^2/(4K {\Lc} \eta)] \comma
\nonumber
\\
 \Theta(u)   & = & \Theta_{Jb}(u) \, \exp [i \pi m_2 (u-K)^2/(4K {\Lc} \eta)] \comma \ea
$H_{Jb}(u), \Theta_{Jb}(u)$ being the usual Jacobi theta  functions (eqn. 15.1.5 of
\cite{Baxter82}). These modified functions are periodic of period $2 {\Lc} \eta$. The zero-field
eight-vertex model Boltzmann weights are then given by (I.8) and (II.6.1):
\ba  \label{8Vweights}
a &  = & \rho \, \Theta(-2\eta) \Theta(\eta-v) H(\eta + v) \comma \nonumber \\
b &  = & -\rho \, \Theta(-2\eta) H(\eta-v) \Theta(\eta + v) \comma \nonumber \\
c &  = &  -\rho \, H(-2\eta) \Theta(\eta-v) \Theta(\eta + v) \comma  \\
d &  = &  \rho \, H(-2\eta) H(\eta-v) H(\eta + v) \period \nonumber \ea

One also uses the functions
\be h(u) = - h(-u) = H(u) \Theta(-u) \comma \ee

\be \phi (u) \eq [\rho \,\Theta(0) h(u) ]^{\L} \comma \ee
which satisfy
\be h(u + \Lc \eta ) = (-1)^{m_1 (m_2 + 1)} \, h(u) \sep \phi(-u) = \phi(u) \comma \ee
\ba \label{hperiods}
h(u + \iota K' ) & = & - e^{-\iota \pi m_1 (2 u+iK')/L \eta } \, h(u) \comma \nonumber \\
h(u + 2 K) & = & - e^{2 \iota \pi m_2 (u+K) /L \eta} \, h(u) \period \ea

This work was re-derived by  Takhtadzhan and  Faddeev (\cite{TakFadd79}) using the
``Quantum Inverse Scattering Method'' (QISM). Then in 1982 the author presented the functional
relation method in sections (10.5) and (10.6) of his book\cite{Baxter82}. An explicit
construction (for ${\L}$ even) of the matrix $\tilde{Q}(v)$ for the eight-vertex model is
given in section 6  of \cite{Baxter73a}, and in section (10.5) of \cite{Baxter82}.

The notation in \cite{Baxter82} is slightly different from that in I, II, III and (\cite{TakFadd79}).
The restriction (\ref{rootofunity}) is not made, $H(u), \Theta(u)$ are the standard theta functions,
the elliptic integrals $K, K'$ are written as $I, I'$, and if we write $\lambda, v$ therein as
$\lambda_B, v_B$, then\footnote{This still leaves $d$ with
a  different sign in \cite{Baxter82} from that in I, II, III, but since vertices 7 and 8 are sinks
and sources of arrows, changing the sign of $d$ does not affect the partition function or the
eigenvalues of the transfer matrix.}
 
\be \label{convert} 
\lambda_B = 2i (K - \eta  ) \sep v_B = 2 i (v -  K) \period \ee

As always, there are  errors and inconsistencies. This is a good opportunity to correct two
of them.

One is a simple but significant typographical error: $j+1$ in eqn. (10.5.8) should be $j-1$, so that
it should read
\be s_j = s + \lambda (\sigma_1 + \cdots + \sigma_{j-1} ) \period
\ee
Equation (10.5.21)  is then consistent with (I.78).\footnote{Apart from $\pm i$ factors that
presumably arise because of the negation of $d$.}

The other is an omission (or at least an over-simplification) by the author in the
Bethe ansatz derivation of eigenvectors in papers I - III. In (7) of
\cite{Baxter71}  each eigenvalue $Q(v)$ of the matrix  $\tilde{Q}(v)$ is taken to be simply a
product of elliptic theta functions. This is corrected in equation (6.10) of \cite{Baxter72},
and in  (10.6.8) of \cite{Baxter82}, where an exponential factor is also included, so that
\be \label{defQB}
Q_B(v) = e^{2 i \tau v} \prod_{j=1}^{\L /2} H_{Jb}(v-u_j) \Theta_{Jb}(v - u_j) \comma \ee
where the suffix $B$ is inserted to distinguish this function from that of III, $\tau$ and $u_1,
\ldots ,u_{\L /2}$ satisfy
\be
\tau = \pi (\hat{s} - 1+ \L  + 4 p' )/8K \comma \ee
\be \label{sumus}
u_1 + \cdots + u_{\L /2}  = (\hat{r} \hat{s} -1 + 4p) K/2 - \iota (\hat{s}-1+\L +4 p') K'/4
\period \ee 
Here $p, p'$ are integers, and $\hat{r} = \pm 1$ is the eigenvalue of the  operator $R$
that reverses all arrows (or spins), and
$\hat{s} = \pm 1$  depending on whether the
number of down arrows is even or odd (and $S$ is the diagonal matrix with  entries $\hat{s}$).
We have used the notation of (10.6.7) - (10.6.8) of 
\cite{Baxter82},  except that we have converted from the $v = v_B$ therein to the present $v,
u_j$ (which are those of papers I - III) by (\ref{convert}).

For $\tilde{n} = \L/2$, the eigenvalue equation (III.1.21) should be the same as  (10.6.1) of 
\cite{Baxter82}. After converting to the notation of III, we find that it is the same, and is
consistent with the other equations (III.1.1) -  (III.1.23),  if extra
$\omega$ factors are included in (III.1.14), (III.1.21), (III.1.23) to make them become
\be \label{defPsi}
\Psi \eq \sum_{l = 1}^{\Lc} \sum_{\bf X} \, \omega^{l} \, f(l| x_1, \ldots x_{\tilde{n}}) \,
\psi(l_1,
\ldots , l_{\L + 1} ) \comma \ee
\be \label{8Veig}
\Lambda \eq  \omega\, \phi(v-\eta) \prod_{j=1}^{\tilde{n}} \frac{h(v-u_j+2 \eta)}{h(v-u_j)} +
 \omega^{-1} \phi(v+\eta) \prod_{j=1}^{\tilde{n}} \frac{h(v-u_j-2 \eta)}{h(v-u_j)} \comma \ee
\be \label{8Vks}
\omega^{-2} \, e^{i {\L} k_j} \eq  - \prod_{m=1}^{\tilde{n}} h(u_j-u_m+2 \eta)/
h(u_j-u_m-2 \eta)  \period \ee
Here \be \label{omegam}
\omega \eq e^{2 \pi i {\tilde{m}}/\Lc} \comma \ee
where the integer $\tilde{m}$ is given by 
\be \label{defmpwr}
2 \, \tilde{m} \eq m_1( \hat{s}-1+\L + 4 p') + m_2 (\hat{r} \hat{s} -1 + 4 p) \period \ee

The other equations amongst (III.1.1) -  (III.1.22) remain unaffected, in particular $e^{\iota
k_j}$ is defined by (III.1.17):  
\be \label{eik8V}
e^{i k_j} = h(u_j + \eta)/h(u_j - \eta) \period \ee
We can still define a function $Q(v)$  by (III.1.24):
\be \label{Q8vdefn}
Q(v) = \prod_{j=1}^{\tilde{n}} h(v-u_j) \comma \ee
but it differs from
$Q_B(v)$  via the exponential factors in (\ref{HvHJb}) and (\ref{defQB}).  Then, exhibiting the
dependence of the transfer matrix eigenvalue $\Lambda$ on $v$, (\ref{8Veig}) can be written as
\be \label{LamQv8V}
\Lambda(v) Q(v) \eq \omega \phi(v-\eta) Q(v+2 \eta) + \omega^{-1} \phi(v+\eta) Q(v- 2 \eta)
\comma \ee
which equation replaces (III.1.25).

These are in fact the correct equations, not just for $\tilde{n} = \L/2$, but for all
$\tilde{n}$ satisfying (\ref{restrnt}).\footnote{Except that the restrictions (\ref{sumus}),
 (\ref{defmpwr}) apply only for $\tilde{n} = \L/2$, which can be regarded as the generic case, and
not necessarily even then if complete strings are present: we  return to this point later in the
section.} We refer to  equations (III.1.1) - (III.1.25), with  the replacements (\ref{defPsi}) -
(\ref{LamQv8V}), as the {\em corrected eight vertex Bethe ansatz equations}.

 There was an omission in the derivation in paper III.  
 Equation (III.2.2) is correct as written, but in (III.3.1)  the author should have
 allowed the  more general ansatz of including a factor
$\omega^l$ in the rhs, where $\omega^{\Lc} = 1$. This is equivalent to associating this factor
with $f(l| x_1, \ldots x_{\tilde{n}})$ and  multiplying
the first term on the rhs of (III.2.2) by $\omega$, the second by $\omega^{-1}$, and (via the
``wanted terms'' and the ``unwanted boundary terms'' of section 3 of III) to the introduction of
the $\omega$ factors in (\ref{defPsi}) - (\ref{LamQv8V}).

Precisely these $\omega$ factors are included in the $\tilde{n} = 0$ equations (II.5.6) and
(II.5.7), where $g_1 = \phi(v+\eta)$ and $g_2 = \phi(v-\eta)$. They are also included in the
work of Takhtadzhan and Faddeev \cite{TakFadd79}.\footnote{Our $\omega$ is $e^{-2 \pi \iota
m/Q}$ in the notation of Takhtadzhan and Faddeev, and it seems that the $l$ of our
equation (\ref{defPsi}) must be the $-l$ of their equation (5.23).}

As Takhtadzhan and  Faddeev comment,\cite[after (5.26)]{TakFadd79} the original equations of
paper III appear to apply only for the case when
$\tilde{m} = 0$. However, it is  better than that. One can verify, for all 
values of $\tilde{n}$ satisfying (\ref{restrnt}), using (\ref{hperiods}), that the
corrected eight vertex Bethe ansatz equations and (\ref{defmpwr})
 are unaffected (apart
from normalization factors that merely renormalize the eigenvector $\Psi$), by the
following simultaneous substitutions:
\bd
u_1 \rightarrow u_1 + \iota K' \sep \omega \rightarrow e^{-4 \iota \pi m_1 /\Lc} \omega
\comma \ed
\bd
\tilde{m} \rightarrow \tilde{m} - 2 m_1 \sep p' \rightarrow p' -1 \period \ed

Similarly, if
$u_1$ is incremented by $2K$, then $\tilde{m}$, $p$ are incremented by $2 m_2$, $1$,
respectively.

The same remarks apply if $u_1$ is replaced by any of the $u_1, \ldots , u_{\tilde{n}}$.

We can use this freedom, which is simply the choice of  period parallelograms for the zeros
$u_1, \ldots , u_{\tilde{n}}$, to increment 
$\tilde{m}$ by any even integer. 
If $\Lc$ is odd, this means that we can construct any choice of $\omega$ by such shifts; while
if $\Lc$ is even, we can construct half the choices. So the 1973 papers do cover
approximately three-quarters of the cases!

More recently, related equations for the eigenvectors of the eight-vertex model have been
studied by Felder and Varchenko\cite{Felder96}, and by Deguchi \cite{Deguchi01b}.

\subsubsection{Alternative form of Bethe's equations}

From (\ref{hperiods}) and (\ref{Q8vdefn}), remembering that $\L$ is even,
\ba
Q(v+ \hat{u} + \iota K') & = &  (-1)^{\tilde{n}} \, e^{- \iota \pi    
  \tilde{n} m_1 ( 2 v+\iota K') /\Lc \eta } \, Q(v + \hat{u} ) \comma \nonumber \\
Q(v+ \hat{u} +2 K) & = &  (-1)^{\tilde{n}} \, e^{2 \iota   \pi  \tilde{n} m_2 
(v  +   K) /\Lc \eta } \, Q(v + \hat{u}) \comma \nonumber \\
\phi (v + \iota K' ) & = &  e^{-\iota  \pi  \L  m_1  (2 v+ \iota K')/\Lc \eta } \, \phi(v)
\comma   \\
\phi (v + 2 K) & = &  e^{2 \iota \pi \L  m_2  (v+K) /\Lc \eta} \, \phi(v) \comma \nonumber
\ea where
\be \hat{u} \eq (u_1 + \cdots + u_{\tilde{n}} )/ \tilde{n} \period \ee

\noindent From (\ref{8Vweights}), or from (\ref{restrnt}) and (\ref{LamQv8V}), the function
$\Lambda (v)$ satisfies the same quasi-periodicity relations as those above for  $\phi (v)$.

The functions $Q(v), \phi(v), \Lambda(v)$ are all entire. Using the $v \rightarrow v + 2 K$
quasi-periodicities above, it follows that there exist coefficients $q_j, f_j, t_j$ such that
\ba
Q(v) & = &  e^{\iota  \pi \tilde{n} m_2 (v-\hat{u})^2/2 K \Lc \eta }\, \sum_j (-1)^j 
q_j \, e^{\iota \pi j' (v - \hat{u})/ K} \comma \nonumber \\
\phi(v) & = &  e^{\iota  \pi \L m_2   v^2/2 K \Lc \eta }\, \sum_j
f_j \, e^{\iota \pi j v/ K} \comma  \\
\Lambda(v) & = &  e^{\iota \pi \L  m_2   v^2/2 K \Lc \eta }\, \sum_j
t_j \, e^{\iota \pi j v/ K} \comma \nonumber \ea
where $j$ takes all positive and negative integer values and 
$j' = j $  if $\tilde{n}$ is even, while  $j' = j - \!$ {\footnotesize{1/2}} if $\tilde{n}$ is
odd. 

Now using the $v \rightarrow v + i K'$
quasi-periodicities, we find that the coefficients in these series must satisfy
\be \label{q8vperds}
 q_{j+\tilde{n}} \eq   e^{ -  \pi (2 j' + \tilde{n}) K'/2K}
\,  q_j \comma \ee
\bd
f_{j+\L} = e^{-\pi (2 j + \L) K'/2 K} \, f_j \sep t_{j+\L} = e^{-\pi (2 j + \L) K'/2 K} \, t_j
\period
\ed

If $q_0, \ldots q_{\tilde{n}-1}$ are known, then the simple periodicity relation
(\ref{q8vperds}) determines all the other
$q_j$. Similarly, all the $t_j$ are determined by $t_0, \ldots ,t_{\L-1}$, and the 
known $f_j$  by $f_0, \ldots ,f_{\L-1}$.

These series are convergent for all
finite $v$. Substituting them into (\ref{LamQv8V}) and equating coefficients, we obtain
\be \label{BA8V2}
\sum_{m =  -\infty}^{\infty} B_{j,m} \, q_m \eq 0 \ee
for all integers $j$, where
\bd
B_{j,m} \eq - t_{j-m} +  e^{\iota  \pi m_2 (8 \tilde{n}  - \L )  \eta/2 \Lc K}
   \, [ \tilde{\omega}f_{j - m + m_2 \nu'  }  + {\tilde{\omega}}^{-1} \, f_{j - m - m_2 \nu'  }
] \comma \ed
where 
\bd
{\tilde{\omega}} \eq \omega \, e^{\iota \pi  (m + 2m' - j ) \eta /K} \,
 e^{-2 \iota \pi \tilde{n} m_2  \hat{u} / \Lc K} \ed 
and  $m' = m $  if $\tilde{n}$ is even, $m' = m -  ${\footnotesize{1/2}}  if $\tilde{n}$
is odd. 

Define
\be  \overline{q}_j \eq  e^{ \pi {j'}^2 K'/2 \tilde{n} K } \, q_j \comma \ee
\be \overline{B}_{j,m} \eq (-1)^{j+m} \, e^{\pi {j'}^2  K'/2 (\L + \tilde{n}) K } \, 
e^{- \pi {m'}^2  K'/2 \tilde{n} K } \, B_{j,m} \period \ee

Then (\ref{BA8V2}) becomes
\be \label{BA8V2a}
\sum_{m = -\infty}^{\infty} (-1)^m \, \overline{B}_{j,m} \, \overline{q}_m \eq 0 \ee
and $\overline{q}_m$, $\overline{B}_{j,m}$ satisfy the periodicity relations
\be  \label{qBpers}
\overline{q}_{m+\tilde{n}} = \overline{q}_m \sep  \overline{B}_{j+\L
+\tilde{n},m+\tilde{n}} =  \overline{B}_{j,m} \; \; \; \; \; \forall \; \;  m,j \period \ee

We can therefore define discrete Fourier transforms $\hat{q}_{\alpha}$,
$\hat{B}_{j,\alpha}$ such that
\be
\overline{q}_m \eq \sum_{\alpha = 0}^{\tilde{n}-1} e^{ 2 \iota \pi  \alpha m/\tilde{n}} \,
\hat{q}_{\alpha} \sep
\hat{B}_{j,\alpha} \eq \sum_{m = -\infty}^{\infty} e^{2 \iota \pi  \alpha m/\tilde{n}} \,
\overline{B}_{j,m} \comma \ee
$m , \alpha $ being integers in the ranges $-\infty < m < \infty$ and $0 \leq \alpha <
\tilde{n}$. Then (\ref{BA8V2a}) becomes 
\be \label{BA8V2b}
\sum_{\alpha = 0}^{\tilde{n} - 1} \hat{B}_{j,\alpha} \, \hat{q}_{\alpha} \eq 0 \period \ee

\noindent Here $j$ can take any integer value, but from (\ref{qBpers})
\be
\hat{B}_{j+\L + \tilde{n},\alpha} \eq  \hat{B}_{j,\alpha} \comma \ee
so there is no loss of information in restricting $j$ in (\ref{BA8V2b}) to lie in the range
\bd 0 \leq j < \L + \tilde{n} \period \ed

Hence (\ref{BA8V2b}) can be regarded as a set of $\L + \tilde{n}$ homogeneous 
linear equations for the $\tilde{n}$ unknowns $\hat{q}_{\alpha}$. It is also linear in the
coefficients $t_j$, and all of these can be determined immediately and linearly from
(\ref{q8vperds}) if we know $t_0, \ldots ,t_{\L -1}$. Thus $t_0, \ldots ,t_{\L -1}$ play the
role of generalized eigenvalues, as $t_0, \ldots , t_{\L}$ do for the six-vertex model in 
(\ref{bqeqn}), (\ref{BAalt2}). Since only the ratios of the $\hat{q}_{\alpha}$ enter the
equation, we have $\L + \tilde{n}$ equations for a total of $\L + \tilde{n} - 1 $ unknowns.
Unlike the six-vertex model, this set is over-determined, presumably because of the
quasi-periodicity constraints satisified by the elliptic functions. They must of course
have solutions.

Many of the remarks made following equation (\ref{LamQ}) for
the six-vertex model extend to the eight-vertex model. One can construct an $ \L + \tilde{n}$ by
$\tilde{n}$ matrix $\bf B$ with elements $\hat{B}_{j,\alpha}$, and write (\ref{BA8V2b})
as  ${\bf B \hat{q}} = {\bf 0}$. Thus ${\bf B}$ must have rank at most $\tilde{n} - 1$, and   
$\bf \hat{q}$ is the column null vector of $\bf B$.\footnote{It is a compication that $\bf B$
depends on $\hat{u}$. This disappears if $m_2 = 0$: possibly it can be removed for
other values by using the periodicity in integer multiples of $\eta$ (rather than $2K$) as the
basis for the discrete Fourier transforms.} If the eigenvalue
$\Lambda(v)$ is degenerate (for all $v$), then the rank of $\bf B$ will be less than $\tilde{n}
- 1$. There will then  be more than one solution for $\bf \hat{q}$, and hence of  (\ref{LamQv8V})
for $Q(v)$. One can expect such behaviour in the situation that we shall now discuss, i.e.  when
$Q(v)$ contains one or more complete strings.

\subsubsection{Similarities to the six-vertex model: strings}

The eight-vertex model is a generalization of the zero-field six-vertex model, and many of the
remarks we have made about the six-vertex model continue to apply. There are still very
special eigenvectors with eigenvalues of the form (\ref{egspec}), namely the 
eigenvectors discussed in papers I and II, corresponding to $\tilde{n} = 0$ in (\ref{8Veig}) .
For $\L$ even, explicit expressions for the matrix
$\tilde{Q}$ are given in \cite{Baxter72} and in section 10.5 of \cite{Baxter82}. From these it
follows that each eigenvalue $Q(v)$ is a product of $\L /2$ elliptic $h(u)$ functions, so the
only way we can get these simple eigenvalues is for all the $u_1, \ldots u_{\L/2}$ to be grouped
into complete strings. These are of length $M$, where
\ba
M  & = & \Lc /2 \; \; {\rm if} \; \; \Lc \; \; {\rm is \; \; even } \comma \nonumber \\
 & = & \Lc  \; \; {\rm if} \; \; \Lc \; \; {\rm is \; \; odd } \period \ea
Each string consists of $M$ zeros, say $u_1, \ldots ,u_M$, differing sequentially by $2
\eta$, i.e. for $j=1, \ldots ,M$
\be \label{typstr}u_{j+1} \eq u_j + 2 \eta   \comma \ee
interpreting $u_{M+1}$ as $u_1$, modulo $2 M \eta$.

In fact any eigenvector and eigenvalue given by (III.1.1) - (III.1.23)\footnote{including 
the modifications (\ref{defPsi}) - (\ref{8Vks})} with $\tilde{n} \neq \L /2$, must also be
present  in the case $ \tilde{n} = \L /2$,
differing from it only in the subtraction or addition of complete strings. The  eigenvectors
for  $\tilde{n} \neq \L /2$ lie in the eigenspace of any other allowed value of
$\tilde{n}$ which is closer to (or equal to) $\L /2$, at least for $\tilde{n} \leq \L /2$.

 For {\em any} values of
$\Lc , m_1, m_2$,  the condition (\ref{restrnt}) is satisfied by $\tilde{n}
= \L /2$, so this is the generic case. The eigenvalue equations (\ref{8Veig}) - (\ref{eik8V})
then apply for all $\eta$. The eigenvector equations, notably (\ref{defPsi}), depend on $\eta$
satisfying (\ref{rootofunity}), but one can approach arbitrarily close to any desired value by
taking  $\Lc , m_1, m_2$ sufficiently large.\footnote{The sum over $l$ in  (\ref{defPsi}) is a
discrete Fourier transform: for general values of $\eta$ it may be appropriate to replace it by
a continuous one.}

Other values of $\tilde{n}$ only occur if (\ref{rootofunity})
is satisfied for $\Lc$ not greater than $\L$.

Again, there are technical difficulties  about handling Bethe's equations when there are
complete strings. The remarks of section 4 extend from the six-vertex to the eight-vertex model.
One should replace (\ref{8Vks}) by the two equations

\be \label{BA8V} \omega^{-2} e^{i {\L} k_j} \eq   \prod_{m=1, m \neq j}^{\tilde{n}}
(- t_{m,j} /t_{j,m}) \comma \ee
 
\be \label{th} t_{j,m}  \, h(u_j-u_m+2 \eta ) = t_{m,j} \, h(u_m-u_j+2 \eta )  \comma \ee
and write (III.1.20) as 
\be \label{Anorm}
A(P) \eq \epsilon_P  \; C^{-1} \prod_{1 \leq j < m \leq {\tilde{n}}} t_{Pm,Pj} \comma
\ee where the renormalization factor $C$ is the same for all permutations $P$ and is the
product of selected
$t_{j,m}$ factors, one from each string, for which $ h(u_m-u_j+2 \eta )$ vanishes.

If $u_j$ does not belong to any string, then in (\ref{BA8V}) we can take $t_{j,m} = h(u_m-u_j+2
\eta ) $. The contribution to the rhs from the $u_m$ that do lie within
strings cancels out, leaving a reduced equation where $j,m$ only take non-string values.
If $u_j$ does belong to a string, say to (\ref{typstr}), then $t_{12}, t_{23}, \ldots ,
t_{M,1}$ vanish but their ratios remain finite. If one fixes one of the $u_j$ within the
string, then the rest are determined, the $e^{\iota k_j}$ are given by (\ref{eik8V}), and the
ratios of $t_{12}, t_{23},
\ldots , t_{M,1}$ are determined by (\ref{BA8V}) for $j = 1, \ldots ,M$. From (\ref{Anorm}),
some of the  coefficients $A(P)$ involve $t_{12}, t_{23}, \ldots , t_{M,1}$ only via these
ratios, so are finite and non-zero. The other $A(P)$ are  of linear or higher order in 
$t_{12}, t_{23}, \ldots , t_{M,1}$, so vanish.

Again, one is free to vary each of the string centres at will.  As one varies these parameters, 
and the
disposable parameters $s, t$ in III, the eigenvector $\Psi$ will trace out a surface $\cal S$ in
the eigenspace of the eigenvalue $\Lambda$. If the eigenvalue is unique, these variations will
merely change the normalization. If it is degenerate, $\cal S$ will lie in the eigenspace
appropriate to this value of $\tilde{n}$, and we expect the vectors on $\cal S$ to span this
eigenspace. 

Since the explicit construction of $\tilde{Q}(v)$ given in section 10.6 of \cite{Baxter82}
gives $\tilde{n} = \L /2$, we expect this to be the generic case, giving all eigenvalues and a
complete set of eigenvectors of $T(v)$.
For $\tilde{n}$ satisfying (\ref{restrnt}), but not equal to $\L /2$, one expects to  only
observe the degenerate eigenvalues, and to obtain only a sub-space of the eigenspace of each.
Recent numerical results support these expectations.\cite{Fabricius}

Of course, one may have particular reasons for fixing the string centres at particular values,
as Fabricius and McCoy did for the six-vertex model.\cite{DegFabMcCoy01} - \cite{FabMcCoy01c}
An obvious choice (for $\tilde{n} = \L /2$) is to fix them so that $Q(v)$ is the eigenvalue of
the matrix $\tilde{Q}(v)$ constructed in section 6 of \cite{Baxter72} and in section (10.5) of
\cite{Baxter82}. Equivalently, one can require that they be fixed to their limiting values
(again, for  $\tilde{n} = \L /2$) as $\eta$ approaches the ``root of unity'' value 
(\ref{rootofunity}). However, these considerations lie outside the Bethe ansatz for a fixed
value of $\eta$. The Bethe ansatz is complete without them: the arbitrariness in the string
centres (and in $s$, $t$) is a reflection of the degeneracy of the eigenvalues of the tranfer
matrix, and the resulting non-uniqueness of the eigenvectors.

\subsubsection{The six-vertex model limit}

In the limit when the elliptic modulus $k$ (or the nome $q$) goes to 0, 1 or $\infty$, the
elliptic functions become trigonometric functions and the eight-vertex model becomes the
zero-field six-vertex model. Much of the working of this section can be adapted at once to the
six-vertex model, except that some of the $u_1, \ldots ,u_{\L/2}$ may become infinite. In this
way the resulting six-vertex model function $Q(v)$ can have any number $n \leq \L/2$ of finite
zeros, and the $\omega$ factors in (\ref{8Veig}) can be related to those in (\ref{eig2b}).
It should be possible to obtain all the six-model eigenvalues from the those of the
eight-vertex model by taking such a limit.

There is a problem with the eigenvectors. For $n \neq \L/2$, the zero-field six-vertex model
eigenvalues occur in degenerate pairs, one in the sub-space with $n$ down arrows, the other in
the arrow-inverted sub-space with $\L-n$ down arrows. Thus two eight-vertex model eigenvalues,
with opposite spin-reversal symmetry, must coalesce. Their sum and difference will then be the
six-vertex model eigenvectors in the two sub-spaces. Only in the $n  = \L/2$ sub-space can one
expect to obtain the six-vertex model eigenvectors directly as limits of those of the
eight-vertex model.

\subsubsection{The sum rule}

The constraint (\ref{sumus}) 
applies to the eight-vertex $Q(v)$ functions obtained by the explicit construction in section (10.6)
of  \cite{Baxter82}, which have $\tilde{n} = \L/2$ zeros. If the eigenvalue $\Lambda)(v)$ is
non-degenerate, i.e. if there are no exact complete strings, then $Q(v)$ is uniquely
defined by Beth's equations, so $\tilde{n}$ will be  equal to $\L/2$, and (\ref{sumus})  will
automatically be satisfied.

However, if there are strings, even if  $\tilde{n} = \L/2$, then one can shift the string centres
arbitrarily and Bethe's equations will be unaffected. The resulting eigenvector will  not necessarily
be an eigenvector of $R$ and $S$, and (\ref{sumus}) will not in general be satisfied,  but it will
nevertheless be a valid eigenvector of the transfer matrix $T(v)$.

So we conclude that, for all $\eta$ satisfying (\ref{rootofunity}), a complete set of $2^{\L}$
eigenvalues and eigenvectors can be obtained by taking $\tilde{n} = \L/2$ and observing the
constraint (\ref{sumus}) (though even then the eigenvectors will not necessarily also be eigenvectors
of $R$ and $S$ for all values of the string centres and the disposable parameters $s$ and
$t$ \cite{Fabricius}). Further eigenvectors can be obtained by abandoning these constraints, while 
of course retaining (\ref{restrnt}), but these eigenvectors will lie in the eigenspaces obtained
with the constraints, so do not extend these eigenspaces.

These remarks extend to the six-vertex model limit. In the $n = \L /2$ sub-space, non-degenerate
eigenvalues must satisfy  the analogue of (\ref{sumus}), which , using (\ref{convert}), is 
\be \label{sixvconstr}
v_1 + \cdots + v_{\L/2} \eq  \iota \pi (r s - 1 + \L + 4 p )  /2 \period \ee
These  $v_1, \ldots , v_{\L /2}$  are those of sections 1 through 6, 
$p$ is an arbitrary integer, $r$ the eigenvalue ($\pm 1$) of the spin reversal operator, 
and $s = (-1)^{\L /2}$  the arrow parity of this state, with $\L /2$ down arrows. This is
 equation (17) of \cite{DoikouNep98}, and equation
(1.44) of \cite{FabMcCoy01c}. Again, if the eigenvalue is degenerate and  $v_1, \ldots , v_{\L /2}$
contain one or mpore complete strings, then (\ref{sixvconstr}) will not necessarily be satisfied. I
am indebted to Barry McCoy and Klaus Fabricius for  correspondence on this and related matters
concerning the Bethe ansatz.

\subsubsection{Other possible difficulties: bound pairs}

Because of the double periodicities of elliptic functions, the technical problems in
the six-vertex model associated with zeros $v_1, \ldots ,v_n$ going to infinity cannot occur in
the eight-vertex model: the corresponding $u_1, \ldots , u_{\tilde{n}}$ can be restricted to a
period parallelogram. There appears to be no ``beyond the equator'' problem.  All the states 
are accounted for by taking $\tilde{n} = \L/2$.  The equations may well have solutions for
$\tilde{n} > \L/2$: this would correspond to adding complete strings to $Q(v)$.

One could still have bound pairs analogous to (\ref{boundpair}), when $-u_1 = u_2 = \eta$,
\be e^{\iota k_1 } \eq e^{-\iota k_2} \eq 0  \sep e^{\iota (k_1+k_2)} = -1 \comma \ee
and the ratio $t_{21}/t_{12}$  vanishes.
One would expect to handle this in the same way as for  the six-vertex model, using (\ref{BA8V})
for $j = 1, 2$ to calculate $e^{\iota Nk_1} t_{12}/t_{21}$ and $e^{\iota Nk_2}
t_{21}/t_{12}$, then substituting these into the suitably renormalized equations for $A(P)$
and the eigenvector $\Psi$.



Although we have not observed the phenomenon, it is conceivable that two or more of $u_1,
\ldots ,u_{\tilde{n}} $, say $u_1, 
\ldots ,u_p$,  could  coincide  at some arbitrary value.  This would require  either
generalizing the argument of section 6, or first dividing  (III.1.16) by the product
of  $h(u_i-u_j)$ over $1 \leq i < j \leq p$, and then taking the limit where $u_1, 
\ldots ,u_p$ become equal.

\section{Summary}

We have presented the coordinate Bethe ansatz equations with some care, trying to avoid 
(or at least signpost) the
problems that occur when some of the variables are zero or infinity. Perhaps the essential
point of this paper is that for the six-vertex model the Bethe ansatz equations are
(\ref{gexp}) - (\ref{sumks}). It seems that one can always choose the coefficients $A(P)$ to
be given by  (\ref{Asoln}), but this is not necessary if enough of the $s_{ij}$ vanish (as
they do when the $v_j$  are equal and  infinite). All that is necessary for $g$ to be an
eigenvector is that (\ref{gexp}) - (\ref{sumks}) (or their appropriately renormalized forms)
be satisfied, with  $ g \neq 0$.

For the six-vertex model, in section 3 we have discussed the situation that arises when some
of the Bethe zeros $v_j$ are infinite, and how this leads to a reduced Bethe equation
containing $\omega$ factors. In section 4 we show that this is the key to resolving the
``beyond the equator'' problem and to constructing the Bethe eigenvector for $n > \L /2$.
We also show how to cope with the problem of a bound pair, when two of the momenta are
infinite but their sum is finite (equal to an odd integer multiple of $\iota \pi$). 

In section 5 we
look at the problem discussed by Fabricius and McCoy, when there are one or more exact
complete strings. We show that the $v_j$ are no longer uniquely determined, because the
string centres can be chosen at will. Nevertheless, the Bethe ansatz equations are satisfied,
and we show how to construct the (necessarily non-zero) eigenvector by working with
appropriate ratios of the vanishing $t_{ij}$.

In particular, we have found the solutions of the Bethe ansatz corresponding to all the
$v_1, \ldots , v_n$ lying on a single complete string. The Bethe ansatz
equations do not define the string centre (the average of $v_1, \ldots
v_n$). This is to be expected: it is a direct consequence of the eigenvalue
$\Lambda$ being degenerate, which means that there is more than one eigenvector, and
hence more than one solution of the Bethe ansatz.  We show that the ansatz 
can be used to construct a {\em complete} set of 
eigenvectors, spanning the eigenspace.

One should of course go on to study more complicated situations, 
where there are more than
one complete strings, and not all the $v_j$ belong to a string. We 
expect the above methods to
generalize to such cases, but the algebra may well be  complicated.

In all the cases dealt with in sections 3 to 5,  one can always use (\ref{gexp}) 
 (with an appropriate choice of the normalization factor  $C$ in
(\ref{Asoln})) as written, each term in the summand being finite and the sum being finite and
non-zero. The only ``limit'' is that of recognizing that one has the set of rational equations
(\ref{eik}), (\ref{st}) - (\ref{sdef2}) to solve for the variables $e^{v_j}$,
$e^{\iota k_j}$ and $t_{ij}$. Some of these variables may be zero or infinite, while
what one wants is their ratios or some other product of powers, which are finite and non-zero.
These ratios or products can of course themselves be regarded as variables. This is only a
generalization of the usual practice of including the ``point at infinity'' in the domain of
the variables.

In section 6 we touch on another  problem, namely  what happens if two or more of the $v_j$
are equal and finite. (The question of what happens when they are equal and infinite
is different, actually easier to resolve, and is dealt with in section 3.) In general this
case really does seem to demand that one take a limit in the expression (\ref{gexp}) for the
eigenvector $g$, since the terms in the summand are finite and non-zero but cancel one
another in pairs, so that their sum is zero. This is different from and less satisfactory than
the other cases, but we remark that in fact we never encountered this case in our numerical
experiments, and it is not the case discussed in \cite{FaddTak84} - \cite{FabMcCoy01c}. It
is not clear that it ever actually occurs. 

In section 7 we discuss the zero-field eight-vertex model with an even number of columns, and
give the needed corrections to the coordinate Bethe ansatz equations of section 1 of
\cite{Baxter73c}. We indicate how the string and infinite momenta problems can occur also for
this model, and how to resolve them.

If one wants to make a specific choice of the string centres, particularly if one wants
to ensure continuity as  $\lambda$ or $\eta$ passes through the ``root of unity'' value
(\ref{lamval}) or (\ref{rootofunity}), or (equivalently) if one wants
$Q	(v)$ to be the eigenvalue of the matrix $\tilde{Q(v)}$ of (86) of \cite{Baxter73a}, or
(10.5.31) of \cite{Baxter82}, then clearly one should use the results of Fabricius and
McCoy. They have addressed this problem in a series of well-presented papers, and have
systematically exhibited  the connections to the 
$sl_2$ loop algebras. However, they do use the provocative title ``Bethe's equation
is incomplete ...''.\cite{FabMcCoy01a} If all one wants to do is to diagonalize the
transfer matrix (or the XXZ hamiltonian), obtaining all the eigenvalues, their degeneracies
and eigenspaces,  then it seems that there is no need to look further than the Bethe ansatz.
At least for the cases studied in this paper, the Bethe ansatz is in fact complete.

The alternative forms (\ref{bqeqn}), (\ref{BA8V2}) of Bethe's equations are themselves
generalized eigenvalue equations in which the $n+1$ or $\tilde{n}$ independent 
coefficents $q_j$
are the elements of the eigenvector. They have some advantages over (\ref{st}), (\ref{BA2}),
(\ref{BA8V}), (\ref{th}), being linear in the $q_j$ and the $t_j$. The ``infinite $v_j$''
problem merely corresponds to some of the coefficients $q_j$ vanishing.\footnote{But in
general one still needs to calculate the zeros  $v_j$ or
$u_j$ in order to obtain the eigenvector.} It seems surprising, but while the author did not
believe these equations to be new, he has been unable to find any previous paper where they 
have been written down.

\section*{Appendix A}

\renewcommand{\theequation}{A\arabic{equation}}
\appendix
\setcounter{equation}{0}

 Suppose for the moment that $A(1,\ldots ,n)$ is non-zero.
Let $\alpha_{m,j}$ be $A(P_{m,j})$, where $P_{m,j}$ is the permutation where $j$ is 
removed from its
place in the sequence $1,2, \ldots ,n$ and replaced immediately after $m$, or immediately before
$m+1$. Thus
\bd \alpha_{m,j} = A(1, \ldots , m,j,m+1,\ldots , j-1,j+1,\ldots,n) \; \; {\rm if}
   \; \; m < j \comma \ed
\be \alpha_{m,j} = A(1, \ldots , j-1,j+1,\ldots , m, j, m+1,\ldots,n) \; \; {\rm if}
   \; \; m \geq j \period \ee
In particular,
\bd \alpha_{0,j} = A(j,1,\ldots ,j-1,j+1,\ldots,n) \comma \ed
\bd \alpha_{j-1,j} = \alpha_{j,j} = A(1,\ldots ,n) \comma \ed
\bd  \alpha_{n,j} = A(1,\ldots ,j-1,j+1,\ldots,n,j) \period \ed

Fix $j$ at some value between 1 and $n$. Then there are $n-1$ equations of the set (\ref{aperm})
that involve only the $n$ distinct coefficients $\alpha_{0,j} , \ldots ,\alpha_{n,j}$, namely
\be s_{m,j} \alpha_{m,j} + s_{j,m} \alpha_{m-1,j} \comma \ee
for $m = 1, \ldots , n$, $ m \neq j$. 
Also, from (\ref{acyc}) we find that
\be e^{i {\L} k_j} \alpha_{n,j} \eq \alpha_{0,j} \period \ee.
Together, these give us $n$ linear homogeneous equations in $n$ unknowns. Since at least one of
the unknowns, namely $\alpha_{j-1,j} = \alpha_{j,j} = A(1,\ldots ,n)$, is
non-zero, the determinant of the matrix of coefficients of these $n$ equations must vanish. This
determinant is easily obtained, giving
\be \label{BAapp} e^{i {\L} k_j }  \prod_{ m=1 , m \neq j} ^n s_{j,m} \eq 
(-1)^{n-1} \! \prod_{ m=1 , m \neq j } ^n s_{m,j}  \ee 
and this must hold for all the possible values $1, \ldots, n$ of  $j$.

Permuting the indices $1,\ldots , n$ merely rearranges the $n$ equations (\ref{BAapp}),
so our initial assumption that $A(1,\ldots ,n)$ is non-zero is irrelevant: to derive
(\ref{BAapp}) it is sufficient that  any one of the coefficients $A(p_1, \ldots, p_n )$ be
non-zero. This must be so for $g$ to be a non-zero vector.




\section*{Acknowledgements} The author thanks his colleagues for their help and advice in
preparing this article, notably Murray Batchelor, Vladimir Bazhanov,   Klaus
Fabricius and Barry McCoy.

\end{document}